\documentclass[reprint, aip, jcp, amsmath,amssymb, superscriptaddress]{revtex4-1}

\usepackage{graphicx}
\usepackage{dcolumn}
\usepackage{bm}
\usepackage{color, soul}
\usepackage{enumitem}
\usepackage[normalem]{ulem}
\usepackage{comment}

\begin{document}
\title{Neuromorphic heat transport effects in a molecular junction}
\author{Renai Chen}
\affiliation{Center for Nonlinear Studies and Theoretical Division, Los Alamos National Laboratory, Los Alamos, New Mexico 87545, USA}
\author{Galen T. Craven}
\affiliation{Theoretical Division, Los Alamos National Laboratory, Los Alamos, New Mexico 87545, USA}
\begin{abstract}
Understanding energy transport at the nanoscale is an open and fundamental challenge in the molecular sciences with direct implications for the design of new electronics, computing devices, and materials. 
While nanoscale energy transport under steady-state conditions has been studied extensively, there is much less known about energy transport under time-dependent driving forces, particularly in the far-from-equilibrium regime. In this work, we use nonequilibrium molecular dynamics simulations and stochastic thermodynamics to investigate energy transport in a well-studied nanoscale system---a molecular junction---subjected to a time-periodic temperature gradient. 
The primary observation is that molecular junctions can exhibit heat transport hysteresis,  a phenomenon in which the heat flux through a system depends not only on the instantaneous value of a time-dependent temperature bias but also on the temporal history of that bias. The presented findings illustrate that molecular junctions can exhibit the specific memory effect---heat transport hysteresis---that is essential for the design of thermal neuromorphic computers. This work elucidates a potential pathway toward the realization of such devices.
\end{abstract}

\maketitle

\section{\label{introduction} Introduction}

Discovering and understanding 
the chemical and physical mechanisms governing energy transport at the nanoscale is one of the most important problems
in the molecular sciences. \cite{Cahill2002, Cahill2003, Dhar2008,Dubi2011,Sato2012,Maldovan2013,Segal2016,Ness2016,Ness2017,Nascimento2022}
In nanoscale systems, the complex interplay, coupling, and often competition between multiple energy transport processes  \cite{Li2012, Segal2016,Sabhapandit2012,Lebowitz1959,Lebowitz1967,Lebowitz1971,Lebowitz2008,Lebowitz2012,Nitzan2003thermal,Segal2005prl,Lebowitz2012,Dhar2015,Velizhanin2015,Esposito2016,craven16c,matyushov16c,craven17a,craven17b,craven17e,craven20a,Ochoa2022}
leads to  phenomena that cannot be described using the same principles that are used for macroscale systems.\cite{BonettoFourier2000,Bonetto2004SCR,Segal2009SCR,Chang2008,craven2023a}
Understanding these transport mechanisms
is central to the development of technologies such as molecular electronics, \cite{Ratner2013review, Xiang2016} 
thermal computing devices, \cite{Joulain2016, craven17b}
thermoelectric materials and molecules, \cite{Gehring2021,Russ2016organic, Sothman2015, Reddy2007}
and  metamaterials with advanced performance characteristics. \cite{Li2021transformingmeta,Yang2024metamaterials}
Over the past several decades, significant progress has been made characterizing steady-state electronic and heat transport at the nanoscale both experimentally and theoretically.
However, less is known about how time-dependent driving forces influence the flow of energy at this scale when the system is pushed from the steady state regime into 
time-dependent nonequilibrium states. \cite{Reimann2002,brey1990generalized,hern07a,hern13d,Ford2015,Seifert2015periodictemp,Seifert2016periodiccurrent,Awasthi2021,Portugal2022effective,Volz2022,Weron2022,Ben-Abdallah2017thermalmemristor,Ordonez-Miranda2019thermalmemristor,Krivtsov2020, Ochoa2024, ochoa2025semiclassical} This is the focus area of this work.
The interplay between an external modulation and energy transport mechanisms can give rise to dynamical effects such as electronic hysteresis and
heat transport hysteresis. \cite{Migliore2013,galperin2005hysteresis,craven2023b, craven2024a, Chen2024, craven2025entropy} 
These phenomena have implications for the design of, for example, energy storage devices and neuromorphic computers.

Theoretical methods to describe nanoscale energy transport
have been developed, and the field is well-established and active. \cite{Segal2016,Nitzan2007,Sato2012,Maldovan2013,Leitner2008,Leitner2013,Seifert2015periodictemp,Li2012,Dubi2011,Lim2013,craven16c,He2021,Volz2022, HernandezJPCL2023, sharony2020stochastic, Krivtsov2023}
One common molecular-level set-up that is used to probe energy transport at the nanoscale is a molecular junction---a device consisting of a single molecule that bridges two electrodes. \cite{Reddy2007,Tan2011,Lee2013,Kim2014,Venkataraman2015,Garner2018,Reddy2019nature,Mosso2019,Zimbovskaya2020}
Applying a temperature bias across the junction by modifying the temperatures of the electrodes induces a heat current through the molecular bridge that can be measured using advanced experimental techniques. \cite{Reddy2019nature,Mosso2019}
Previous experimental studies have primarily focused on steady-state thermal transport across molecular junctions in the presence of a static temperature gradient.  Applying time-dependent temperature gradients at the nanoscale is experimentally challenging, although advances in nanoscale thermal modulation---such as ultrafast photothermal heating and electrically-controlled temperature fields---may enable such protocols.
Segal, Nitzan, and H\"{a}nggi analyzed a molecular junction under steady state conditions using simulations and 
found that a transition between ballistic and diffusive heat transport regimes occurred as the length of the molecular junction was varied. \cite{Nitzan2003electron}
The results of single molecule thermal conductance experiments performed almost two decades later are in agreement with their findings \cite{Reddy2019nature,Mosso2019} 
highlighting the important interplay between theory and experiment in this field.

One potential technological application of nanoscale energy transport is in the development of thermal computing devices. \cite{Li2006,Ben-Abdallah2014,Joulain2016,Wang2017thermaldiode,craven17a,craven17b} 
Thermal computing is a hardware computational approach that involves using heat currents instead of electrical currents to perform logical operations. \cite{Li2007logic}
Some of the potential advantages of thermal computing in comparison to classical electronic computing are:
(1) The ability to harness omnipresent thermal energy sources to perform computations, bypassing the need to generate electrical currents and
(2) Because thermal energy is not a detriment in these devices, the density of transistors on a chip could potentially exceed what is possible with electronic transistors while maintaining operational stability. 
One distinct disadvantage of thermal computers is that the slew rate, i.e., the ON-OFF rate, of devices like thermal transistors will be orders of magnitude slower than electronic counterparts. 
However, this limitation on operational speed may be offset by the ability to increase the density of  transistors on a chip.
Overall, it is important to note that thermal computing is an emerging field and so its practical impact on the technological landscape is
not presently clear.
Fields such as thermotronics and phononics, which aim to control heat exchange and energy flow in devices, are closely related to thermal computing,\cite{Li2012, Ben-Abdallah2017thermalmemristor} and have technological uses for thermal management and thermal information processing.

Another emerging computing architecture is neuromorphic computing. 
The field of neuromorphic computing has gained significant attention due to its potential applications in new computing architectures that increase computing efficiency with respect to both energy usage and operational speed. \cite{Van2018organic,markovic2020physics,kudithipudi2025neuromorphic}
Neuromorphic devices increase computing efficiency by co-locating memory and processing units. \cite{Caravelli2018,Sangwan2020neuromorphic,Yuriy2011, HernandezJPCL2023}
Because of this co-location, neuromorphic computers are energy efficient and avoid the von Neumann bottleneck that arises from transferring information back and forth between separate memory and processing units---a limitation of  standard classical computers.
The principal neuromorphic components are memristors (memory resistors) and memcapacitors (memory capacitors),
devices that perform both memory and processing operations.

The physical phenomenon that gives rise to the memory functionality in traditional neuromorphic devices is electronic hysteresis as signified by hysteresis curves in the $I \times V$ plane where $I$ is electronic current and $V$ is voltage bias.
Models of thermal neuromorphic devices have also been developed, 
and these devices exhibit heat transport hysteresis, i.e., hysteresis curves in the $J \times \Delta T$ plane
where $J$ is heat current and $\Delta T$ is temperature bias.
\cite{Ben-Abdallah2017thermalmemristor,Ordonez-Miranda2019thermalmemristor}
Heat transport hysteresis is a phenomenon in which an energy flux generated by a temperature gradient that is oscillating in time is out of phase with the oscillation pattern of the gradient itself. 
This type of hysteresis behavior is not present in systems that operate under steady-state conditions.  
We have previously shown in lattice models that 
nanoscale systems can exhibit heat transport hysteresis. \cite{craven2023b, craven2024a,Chen2024, craven2025entropy}

In this work, we show the presence of neuromorphic heat transport effects in a model molecular junction
subjected to a periodically-modulated temperature gradient.
We extend our previous work in this area \cite{craven2023b, craven2024a,Chen2024, craven2025entropy}
by using stochastic nonequilibrium molecular dynamics simulations to examine alkane molecular junctions and demonstrate the emergence of heat transport hysteresis in these nonlinear systems. 
We analyze the dependence of hysteresis on system parameters such as temperature oscillation frequency and chain length and
provide insights into the microscopic mechanisms governing nonequilibrium energy transport in molecular systems. 
Our theoretical findings elucidate a potential path to realize neuromorphic heat transport effects in molecular junctions by controlling heat flow using time-dependent temperatures.
Molecular junctions with tunable hysteresis characteristics could function as molecular thermal logic or memory elements. Such systems are well-studied under steady-state conditions, and may ultimately be used in thermal neuromorphic architectures that are controlled using temperature fields. This work provides a mechanistic foundation for exploring time-dependent thermal transport and heat transport hysteresis at the nanoscale.

\section{Heat Transport Hysteresis}
\label{sec:model}

\begin{figure*}
	\includegraphics[width=\textwidth]{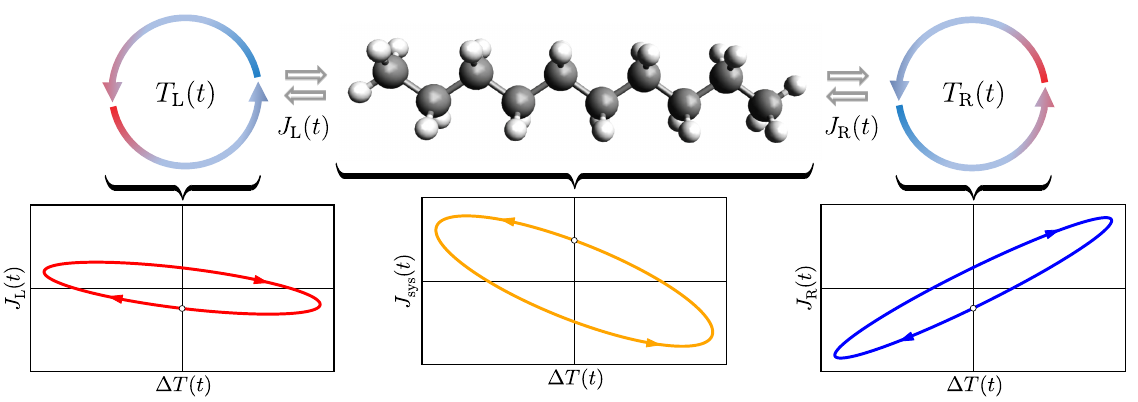}
	\caption{Schematic diagram of an alkane molecular chain connecting two heat baths with time-dependent oscillating temperatures. 
	The temperatures of the left and right baths, $T_\text{L}(t)$ and $T_\text{R}(t)$, are oscillatory as signified by the 
        circular patterns representing each bath.
    The representative hysteresis curves in the bottom row illustrate the concept of hysteresis under thermal driving. The molecular simulation results presented in this paper often deviate from this simple hysteresis curve shape.
    }
	\label{fig:schematic}
\end{figure*}

The first model we consider is an alkane molecular junction consisting of $N$ carbon atoms and $2N+2$ hydrogen atoms that connects two heat baths with 
time-dependent temperatures. 
There are $3N+2$ total atoms in the system
and the value of $N$ defines the effective length of the junction.
See Fig.~\ref{fig:schematic} for a schematic diagram of the system. 
The baths are denoted ``L" for left bath and ``R" for right bath. 
The corresponding time-dependent temperatures $T_\text{L}(t)$ and $T_\text{R}(t)$ 
oscillate at the respective frequencies $\omega_\text{L}$ and $\omega_\text{R}$.
This model is based on the set-ups used in recent experimental measurements of single-molecule thermal conductance under steady-state conditions, \cite{Reddy2019nature,Mosso2019} however note that our investigation is under time-periodic temperature driving.
The primary effect we will show is heat transport hysteresis, which is shown schematically in the bottom row of plots in Fig.~\ref{fig:schematic}.
The schematic hysteresis curves in Fig.~\ref{fig:schematic} illustrate the qualitative concept of thermal hysteresis. As we will show, significantly more complex hysteresis curves can be generated in molecular junctions.

We first examine energy transport in this molecular junction model using molecular dynamics (MD) simulations. The MD simulations are implemented using classical nonequilibrium Langevin dynamics implemented in 
a codebase developed by us.
We use the Optimized Potentials for Liquid Simulations - All-Atom (OPLS-AA) force field, which is a classical force field that
is well-used to understand the properties of organic liquids.
\cite{jorgensen1996development}
With the orientation of the alkane molecule as in Fig.~\ref{fig:schematic},
the equation of motion (EoM) for the left terminal carbon in the chain is 
\begin{equation}
m_\text{C}\,\ddot{\mathbf{r}}_{1}(t)
=
-\nabla_{\mathbf{r}_{1}}U\bigl( \mathbf{r}_{1}, \ldots, \mathbf{r}_{3N+2}\bigr) - m_\text{C}\,\gamma_{\mathrm{L}}\,\dot{\mathbf{r}}_{1}(t) + \boldsymbol{\xi}_{\mathrm{L}}(t),
\end{equation}
and the EoM for the right terminal carbon is
\begin{equation}
m_\text{C}\,\ddot{\mathbf{r}}_{N}(t)
=
-\nabla_{\mathbf{r}_{N}}U\bigl(\mathbf{r}_{1}, \ldots, \mathbf{r}_{3N+2}\bigr) - m_\text{C}\,\gamma_{\mathrm{R}}\,\dot{\mathbf{r}}_{N}(t) + \boldsymbol{\xi}_{\mathrm{R}}(t).
\end{equation}
The EoM for all other (non-terminal) carbons is
\begin{equation}
m_{\text{C}}\,\ddot{\mathbf{r}}_{j}(t) = -\nabla_{\mathbf{r}_{j}}U\bigl( \mathbf{r}_{1}, \ldots, \mathbf{r}_{3N+2}\bigr) : 1<j<N,
\end{equation}
and for all the hydrogen atoms
\begin{equation}
m_{\text{H}}\,\ddot{\mathbf{r}}_{j}(t)
=
-\nabla_{\mathbf{r}_{j}}U\bigl( \mathbf{r}_{1},\ldots, \mathbf{r}_{3N+2}\bigr) : N<j\leq 3N+2.
\end{equation}
In these equations, $m_\text{C}$ is the mass of a carbon atom, $m_\text{H}$ is the mass of a hydrogen atom, $\mathbf{r}_{j} =  \{x_j(t), y_j(t), z_j(t)\}$ is the position of the $j$th atom, $\gamma_\text{L}$ and $\gamma_\text{L}$ are system-bath coupling constants for the left and right baths, and 
$\boldsymbol{\xi}_\text{L}(t) = \{\xi^{(x)}_{\text{L}}(t),\xi^{(y)}_{\text{L}}(t),\xi^{(z)}_{\text{L}}(t)\}$ and $\boldsymbol{\xi}_\text{R}(t) = \{\xi^{(x)}_{\text{R}}(t),\xi^{(y)}_{\text{R}}(t),\xi^{(z)}_{\text{R}}(t)\}$ are stochastic noise terms for the baths that obey the correlations
\begin{equation}
\begin{aligned}
\label{eq:FD_theorem_t}
 \big\langle \boldsymbol{\xi}_\text{L}(t) \boldsymbol{\xi}^\text{T}_\text{L}(t')\big\rangle &= 2 \boldsymbol{\gamma}_\text{L}m_\text{C} k_\text{B} T_\text{L}(t)\delta(t-t') \in \mathbb{R}^{3 \times 3}, \\
 \big\langle \boldsymbol{\xi}_\text{R}(t) \boldsymbol{\xi}^\text{T}_\text{R}(t')\big\rangle &= 2 \boldsymbol{\gamma}_\text{R}m_\text{C} k_\text{B} T_\text{R}(t)\delta(t-t') \in \mathbb{R}^{3 \times 3}, \\
	\big\langle \boldsymbol{\xi}_\text{L}(t) \boldsymbol{\xi}^\text{T}_\text{R}(t')\big\rangle &= \boldsymbol{0} \in \mathbb{R}^{3 \times 3}, \\
 \big\langle \boldsymbol{\xi}_\text{L}(t)\big\rangle &=\boldsymbol{0} \in \mathbb{R}^{3},\\
 \big\langle \boldsymbol{\xi}_\text{R}(t)\big\rangle &=\boldsymbol{0} \in \mathbb{R}^{3},
\end{aligned}
\end{equation}
where $k_\text{B}$ is the Boltzmann constant and $T_\text{L}(t)$ and $T_\text{R}(t)$ are the time-dependent temperatures of the left and right bath, respectively.
The matrices
\begin{equation}
			\boldsymbol{\gamma}_\text{L} =
		\begin{pmatrix}
			\gamma_\text{L} &0&0 \\
			0&\gamma_\text{L} &0 \\
			0&0&\gamma_\text{L} \\
		\end{pmatrix} \quad \text{and} \quad \boldsymbol{\gamma}_\text{R} =
		\begin{pmatrix}
			\gamma_\text{R} &0&0 \\
			0&\gamma_\text{R} &0 \\
			0&0&\gamma_\text{R} \\
		\end{pmatrix}
\end{equation}
are square diagonal matrices of system-bath coupling parameters.
The potential energy surface $U$ is the OPLS-AA force field with an additional harmonic pinning potential on the terminal carbons.
We write this system of equations explicitly here to illustrate that the two terminal carbon atoms are governed by Langevin dynamics
with the baths having time-dependent temperatures.

We will consider the case where the temperatures of each bath take the forms:
\begin{equation}
\begin{aligned}
\label{eq:t-dep_T}
 T_\text{L}(t)   &= T^{(0)}_\text{L} + \Delta T_\text{L} \sin(\omega_\text{L} t),\\[1ex]
 T_\text{R}(t)   &= T^{(0)}_\text{R} + \Delta T_\text{R} \sin(\omega_\text{R} t),
\end{aligned}
\end{equation}
where $\Delta T_\text{L}$ and $\Delta T_\text{R}$ define the amplitude of the oscillations, $\omega_\text{L}$ and $\omega_\text{R}$ are oscillation frequencies as stated before, and $T^{(0)}_\text{L}$ and $T^{(0)}_\text{R}$ are the temperatures of the two baths when there are no oscillations, that is when $(\Delta T_\text{L}$, $\Delta T_\text{R}) \to (0,0)$.
The instantaneous temperature difference between the two baths is 
\begin{equation}
\Delta T(t) =    T_\text{L}(t) - T_\text{R}(t).
\end{equation}
Only cases in which $\omega_\text{L}$ and $\omega_\text{R}$ are commensurate are considered.
Therefore, the system dynamics are periodic with total period $\mathcal{T}$.
Because of this periodicity, the system does not relax to a nonequilibrium steady state, but instead to 
a time-dependent nonequilibrium state (TDNS).
For ease of mathematical exposition, we define an effective temperature parameter 
\begin{equation}
T = \frac{\gamma_\text{L}T^{(0)}_\text{L} + \gamma_\text{R}T^{(0)}_\text{R}}{\gamma_\text{L} + \gamma_\text{R}},
\end{equation}
and an effective system-bath coupling
\begin{equation}
\gamma = \gamma_\text{L} + \gamma_\text{R}.
\end{equation}

The oscillating temperature bias induces time-dependent energy fluxes (heat currents) in the model.
The energy fluxes are separated into three terms:
\begin{enumerate}
  \item $J_\text{sys}$ is the energy flux in/out of the system, i.e., the molecular bridge
  \item $J_\text{L}$ is the energy flux associated with the left bath 
  \item $J_\text{R}$ is the energy flux associated with the right bath
\end{enumerate}
Note that $J_\text{sys} = \frac{d \langle E(t) \rangle}{dt}$ where $E(t)$ is the system energy, and so the system energy flux 
will vanish in the steady-date regime where $E(t)$ is constant. 
Using the stochastic energetics formalism, the average energy fluxes in/out of the baths and the system are \cite{Lebowitz1959,Sekimoto1998,Sabhapandit2012,Dhar2015}:
\begin{align}
\label{eq:heatcurrentbathL}
J_\text{L} (t) &=  m \gamma_\text{L} \big\langle  v_1^2(t) \big\rangle-m \big\langle \xi_\text{L}(t) v_1(t)\big\rangle ,\\[1ex]
\label{eq:heatcurrentbathR}
J_\text{R} (t) &=  m \gamma_\text{R} \big\langle  v_N^2(t) \big\rangle-m \big\langle \xi_\text{R}(t) v_N(t)\big\rangle ,\\[1ex]
\label{eq:heatcurrentbathS}
J_\text{sys} (t) &= \frac{d \big\langle E(t)\big\rangle}{dt},
\end{align}
where $\big\langle  v_1^2(t) \big\rangle$ and $\big\langle  v_N^2(t) \big\rangle$ are the expectation values of the squared velocities of the left and right terminal carbons, respectively, and $\big\langle \xi_\text{L}(t) v_1(t)\big\rangle$ and $\big\langle \xi_\text{R}(t) v_N(t)\big\rangle$ are noise-velocity correlation functions.

The MD simulations are implemented as follows. The alkane junction starts in the minimum energy configuration obtained using 
steepest descent on the MMFF94 force field. \cite{halgren1996merck}
The length of the carbon chain during the simulation is dictated by the minimum energy configuration. When a pinning potential is used, the length of the chain will fluctuate around the length of the minimum energy configuration.
Then, the system is simulated for multiple total periods of oscillation using nonequilibrium Langevin dynamics.
The timestep used in the the simulations was $\Delta t = 0.001$ in AKMA (\AA ngstr\"oms, kilocalories per mole, atomic mass units) units.
Throughout this article, parameters are given in AKMA units with temperature given in units of Kelvin.
The first half of the simulation time is used to  relax the system into the TDNS
with bath temperatures oscillating during this entire procedure.
Once the TDNS is approximately reached,
the energy fluxes are sampled using stochastic thermodynamics methods. \cite{Sekimoto1998, Sabhapandit2012}
Hundreds of thousands to millions of trajectories are typically needed to converge the energy flux calculations, 
and so the procedure is computationally taxing. The energy flux calculations often still have such high noise levels even when averaged over $\sim\!10^7$ trajectories that smoothing methods must be used to elucidate the underlying signals, i.e., the energy fluxes. 

The numerical procedure we use to obtain the energy flux in the baths
is based on work in Ref.~\citenum{sabhapandit2012heat}.
Over the time interval $[t,t+\Delta t]$ the energy change in each of the respective baths is given by
\begin{align}
\label{eq:heat}
 \mathcal{Q}_\text{L}(t) &=  \int_t^{t+\Delta t} \left(m \gamma_\text{L}  v_1^2(t') -m  \xi_\text{L}(t') v_1(t') \right) dt',\\
  \mathcal{Q}_\text{R}(t) &=  \int_t^{t+\Delta t} \left(m \gamma_\text{R}  v_N^2(t') -m  \xi_\text{R}(t') v_N(t') \right) dt',
\end{align}
while the energy change in the system is $\Delta E(t) = E(t+\Delta t) - E(t)$.
In simulation, the expectation values of the energy fluxes are calculated using
\begin{align}
J_\text{L} (t) &=  \sum^n_{i} \frac{\mathcal{Q}_\text{L}^{(i)}(t)}{n \Delta t} ,\\[1ex]
J_\text{R} (t) &=  \sum^n_{i} \frac{\mathcal{Q}_\text{R}^{(i)}(t)}{n \Delta t},\\[1ex]
J_\text{sys} (t) &= \sum^n_{i} \frac{\Delta E^{(i)}(t)}{n \Delta t},
\end{align}
where $n$ is the number of trajectories and the $i$ superscript is an index over each trajectory. 
In all heat transport calculations,  
each cartesian component is analyzed and included in the calculation.

In some of the models examined in this work, particularly for longer alkanes and systems with weak system-bath coupling, after simulating for greater than $\approx\!25\mathcal{T}$ (25 oscillation periods) the energy fluxes had not fully relaxed to the TDNS from the initial conditions. 
Therefore, in these cases, fully relaxing the energy fluxes to the TDNS was prohibitive computationally.
In order to correct for this incomplete relaxation, 
we vertically shifted the energy fluxes. 
We examined the rate of relaxation to the TDNS and observed that this relaxation was occurring over successive periods, albeit approximately asymptotically. We also noted that the general shape of the hysteresis curves was not changing dramatically over successive periods, only the vertical location in the $J \times \Delta T$ plane was changing. We therefore concluded that the
energy shift was appropriate to describe the system dynamics in the TDNS.
The shifting factor $\Delta J_\text{sys}$ was found 
by constraining the time integral over the system energy flux for one period of oscillation to be zero:
\begin{equation}
\int_0^\mathcal{T} J_\text{sys}(t) \, dt  = 0,
\end{equation}
where the equality constraint holds up to some numerical tolerance due to thermal noise and other factors. 
In order to satisfy this constraint, the system energy flux can be written as
\begin{equation}
J_\text{sys}(t) = J'_\text{sys}(t) + \Delta J_\text{sys},
\end{equation}
where $J'_\text{sys}(t)$ is the unshifted simulation data. 
The bath energy fluxes were shifted according to
\begin{equation}
J_\text{L/R}(t) = J'_\text{L/R}(t) - \Delta J_\text{sys} \left(\frac{\gamma_\text{L/R} T^{(0)}_\text{L/R}}{\gamma_\text{L} T^{(0)}_\text{L} + \gamma_\text{R} T^{(0)}_\text{R}}\right),
\end{equation}
where $J'_\text{L/R}(t)$ is the corresponding heat bath energy flux and the factor in the last term on the RHS accounts for the energy partitioning between baths during thermal energy fluctuations. \cite{craven18a1,craven18a2}
Throughout this work we will note when the shifting procedure was performed.

\begin{figure}[]
\includegraphics[width = 8.6cm,clip]{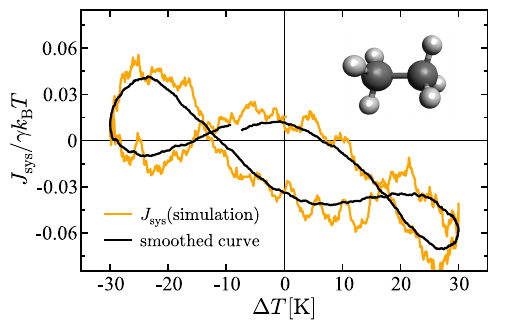}
\caption{\label{fig:ethane_Hys_1}
System energy flux as a function of temperature difference for an ethane molecular junction.
The parameters in the model are $N =2$ (ethane), $\gamma_\text{L} \approx 0.05$, $\gamma_\text{R} \approx 0.05$, $T^{(0)}_\text{L} = 300$, $T^{(0)}_\text{R} = 300$,
$\Delta T_\text{L} = 0$, $\Delta T_\text{R} = 30$, $\omega_\text{L} =0$, $\omega_\text{R} = 5$.
All parameters throughout the article are given in AKMA units.
}
\end{figure}

Figure~\ref{fig:ethane_Hys_1} illustrates the system energy flux for an ethane ($N=2$) molecular bridge connecting two baths, one of which has an oscillating temperature and the other has a  static temperature. 
Both baths have the same baseline temperature $T^{(0)}_\text{L} = T^{(0)}_\text{R}$.
In this figure, and all others throughout, results are shown for one total oscillation period $\mathcal{T}$.
There are two important initial observations: (1) heat transport hysteresis is observed and
(2) a complex hysteresis curve emerges with multiple pinch points where the trajectory of the curve intersects with itself.
This type of curve has not been observed in linear systems with only single-frequency temperature driving, \cite{craven2023b,craven2024a} which highlights that nonlinear effects in the molecular junction can significantly alter heat transport. 
Under steady-state conditions, the heat transport dynamics of a molecular junction can often be well-approximated using a linear models. \cite{Nitzan2003thermal}
The presence of the pinched loops shows that the presence of the oscillating temperature  drives the system into a state that cannot be well described using linear approximations. 

The data shown in Fig.~\ref{fig:ethane_Hys_1} was obtained by averaging over $\sim\!10^6$ trajectories, where each trajectory represents a different realization of the thermal noise.
Even after averaging over this large amount of simulation data, significant thermal noise still persisted in the hysteresis curves making it difficult to distinguish the energy fluxes from the noise. 
Therefore, smoothing was performed.
The specific smoothing procedure we used was to take cumulative average over a time window of sampling points.
Overall, this figure illustrates the emergence of heat transport hysteresis in an alkane model 
and that nonlinear effects may feature prominently when describing the geometry of hysteresis curves in such systems.

\begin{figure}[]
\includegraphics[width = 8.6cm,clip]{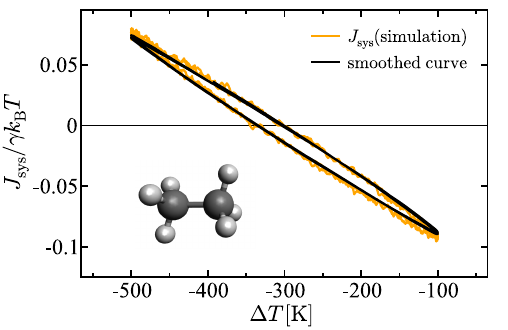}
\caption{\label{fig:ethane_Hys_4}
System energy flux as a function of temperature difference for an ethane molecular junction.
The parameters in the model are $N =2$ (ethane), $\gamma_\text{L} =1$, $\gamma_\text{R} =1 $, $T^{(0)}_\text{L} = 300$, $T^{(0)}_\text{R} = 600$,
$\Delta T_\text{L} = 0$, $\Delta T_\text{R} = 200$, $\omega_\text{L} =0$, $\omega_\text{R} = 20$. The results are shown over two oscillation periods.}
\end{figure}

The pinched hysteresis effects for single-frequency temperature driving are not present in all cases, however. This effects appears to be system specific
and depends on the specific temperature driving frequency and system-bath couplings that are used.
Figure~\ref{fig:ethane_Hys_4} shows the results for an ethane ($N=2$) molecular bridge connecting two baths with stronger system bath couplings, a different baseline temperature gradient, and a faster temperature oscillation frequency than in Fig.~\ref{fig:ethane_Hys_1}.
In this case, the hysteresis presents as a single elliptical loop, similar to our theoretical predictions in Refs.\citenum{craven2023b} and \citenum{craven2024a}
and to what is illustrated schematically in Fig.~\ref{fig:schematic}. Overall, these results illustrate that the existence and shape of the molecular hysteresis curves is affected by a combination of molecular properties, system-bath coupling strengths, and the specific temperature driving protocol that is implemented.

\begin{figure}[]
\includegraphics[width = 8.6cm,clip]{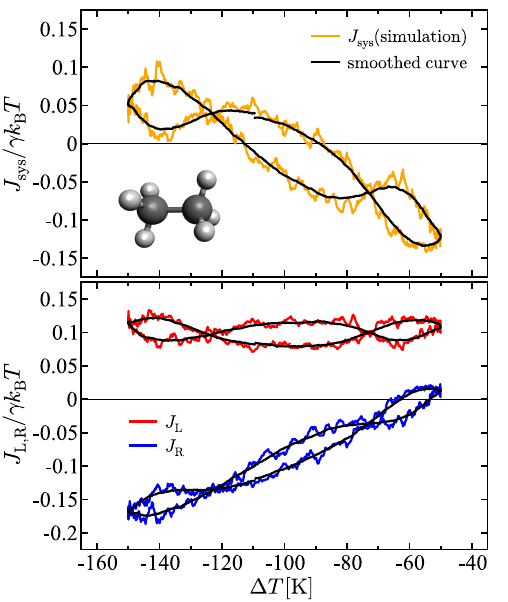}
\caption{\label{fig:Hys_3_ethane}
System energy flux as a function of temperature difference for an ethane molecular junction.
The parameters in the model are $N =2$ (ethane), $\gamma_\text{L} \approx 0.05$, $\gamma_\text{R} \approx 0.05$, $T^{(0)}_\text{L} = 200$, $T^{(0)}_\text{R} = 300$,
$\Delta T_\text{L} = 0$, $\Delta T_\text{R} = 50$, $\omega_\text{L} =0$, $\omega_\text{R} = 5$.
}
\end{figure}

\begin{figure}[]
\includegraphics[width = 8.6cm,clip]{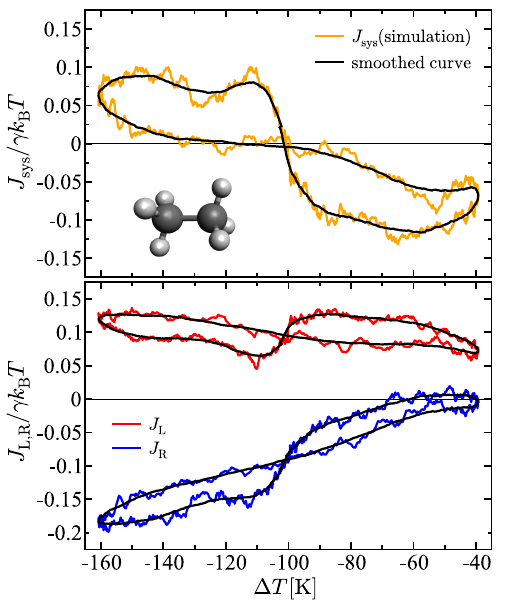}
\caption{\label{fig:ethane_Hys_2}
System energy flux (top) and the energy fluxes of the baths (bottom) as a function of temperature difference for an ethane molecular junction.
The parameters in the model are $N =2$, $\gamma_\text{L} \approx 0.05$, $\gamma_\text{R} \approx 0.05$, $T^{(0)}_\text{L} = 200$, $T^{(0)}_\text{R} = 300$,
$\Delta T_\text{L} = 20$, $\Delta T_\text{R} = 50$, $\omega_\text{L} =10$, $\omega_\text{R} = 5$.
}
\end{figure}

Shown in Fig.~\ref{fig:Hys_3_ethane} are the hysteresis curves for the same setup as in Fig.~\ref{fig:ethane_Hys_1} 
but in this case with a baseline temperature gradient, i.e.,  $T^{(0)}_\text{L} \neq T^{(0)}_\text{R}$. Therefore in this setup in the absence of temperature oscillations there is a static temperature difference between baths, and that temperature difference supports a heat current through the model.
A similar multiple pinched loop pattern (compare with Fig.~\ref{fig:ethane_Hys_1}) is observed for the system energy flux shown in orange in the top panel, signifying that the multi-pinch loop behavior may be a general shape that occurs due to nonlinearity in the ethane molecule force field. 
The energy fluxes for each of the two baths are shown in the bottom panel of Fig.~\ref{fig:ethane_Hys_2}.
Prominent multi-pinch hysteresis loops are also observed for both heat baths.
The left bath is shown in red and the right bath is shown in blue.
Note that the sign convention we use here and throughout is that when energy leaves a bath the energy flux sign is negative while when energy is absorbed by a bath the sign is positive. 
This formalism agrees with the simulation results shown in the bottom panel of Fig.~\ref{fig:Hys_3_ethane}.

\begin{figure}[]
\includegraphics[width = 8.6cm,clip]{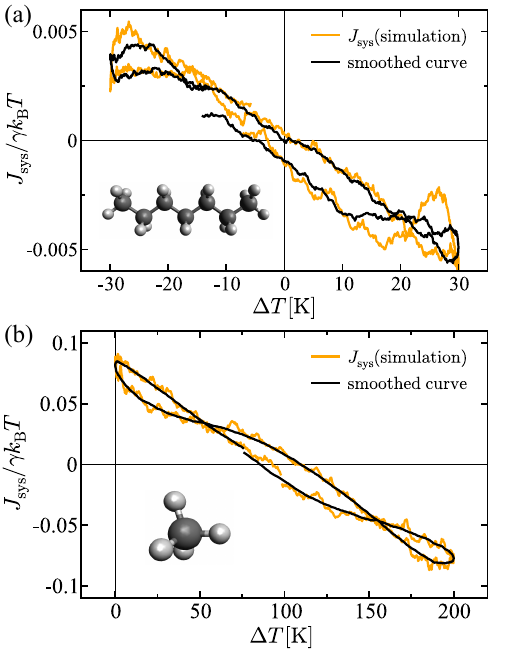}
\caption{\label{fig:septane_Hys_1}
System energy flux as a function of temperature difference for (a) a septane molecular junction and (b) a methane molecular junction.
The parameters in (a) are $N =7$, $\gamma_\text{L} \approx 0.05$, $\gamma_\text{R} \approx 0.05$, $T^{(0)}_\text{L} = 300$, $T^{(0)}_\text{R} = 300$,
$\Delta T_\text{L} = 0$, $\Delta T_\text{R} = 30$, $\omega_\text{L} =0$, $\omega_\text{R} = 5$.
The data in (a) was vertically shifted by a factor $\Delta J_\text{sys}/\gamma k_\text{B} T = 0.014$. 
The parameters in (b) are $N =1$, $\gamma_\text{L} =0.1$, $\gamma_\text{R} =0.1$, $T^{(0)}_\text{L} = 300$, $T^{(0)}_\text{R} = 200$,
$\Delta T_\text{L} = 0$, $\Delta T_\text{R} = 100$, $\omega_\text{L} =0$, $\omega_\text{R} = 10$.
}
\end{figure}

Figure~\ref{fig:septane_Hys_1} (a) illustrates the system energy flux hysteresis curve for the same setup as in Fig.~~\ref{fig:ethane_Hys_1}  except the ethane bridge is replaced with a longer septane ($N=7$) bridge.
In this case, heat transport hysteresis is again observed, but the shape of the curve is altered from the $N=2$ case. 
Notably, the amplitude of the energy flux oscillations is dramatically less than for the shorter $N=2$ system. 
This suggests that shorter molecular chains may be better candidates to observe heat transport hysteresis.
This implies that even when there are more degrees of freedom present in a system 
in which to deposit and store energy (for example, in a longer molecular bridge), energy passes through the larger system (septane) at an increased rate relative to the smaller system (ethane). 
This is evidence for a transition from diffusive to ballistic transport. 
From a geometrical perspective, the prominent two pinched loop that are generated for the $N=2$ case are much less prominent in the $N=7$ case.
However, this pinched behavior can be observed in junctions with other temperature driving protocols and system parameters. Figure~\ref{fig:septane_Hys_1} (b) shows the same type of hysteresis behavior in a methane ($N=1$) junction with different system parameters than Fig.~\ref{fig:septane_Hys_1} (a) and Fig.~~\ref{fig:ethane_Hys_1}. This suggests that the multiple pinch points could be a general behavior in specific physical regimes.

Figure~\ref{fig:ethane_Hys_2} illustrates the system energy flux for an ethane molecular bridge connecting a  hot bath and a cold bath, where the temperatures of both baths are oscillating in time but with different frequencies.
Specifically, the frequencies of the left and right baths are $\omega_\text{L} =10$ and $\omega_\text{R} = 5$ in AKMA units.
A notable  geometric feature arises---a single pinched loop pattern. 
In neuromorphic computing device design, a pinched loop pattern is a required functionality for the most important component, a memristor \cite{chua2014if}, and we observe a similar effect here.
The energy flux has symmetry about the $x$-axis in the sense that approximately half the loop is positive and half the loop is negative. 
This symmetry shows that the system (the molecular bridge) is both acquiring and releasing that same amount of energy on the average over each oscillation cycle thereby sustaining a net current between the hot and cold baths. 
The energy fluxes for each of the two baths are shown in the bottom panel of Fig.~\ref{fig:ethane_Hys_2}.
Prominent single-pinch hysteresis loops are observed for both baths.
The temperature bias in these figures is motivated by two considerations: (1) to drive the system  beyond the linear-response regime and (2) to reflect experimentally accessible temperature differences in nanoscale thermal control platforms.

\begin{figure}[]
\includegraphics[width = 8.6cm,clip]{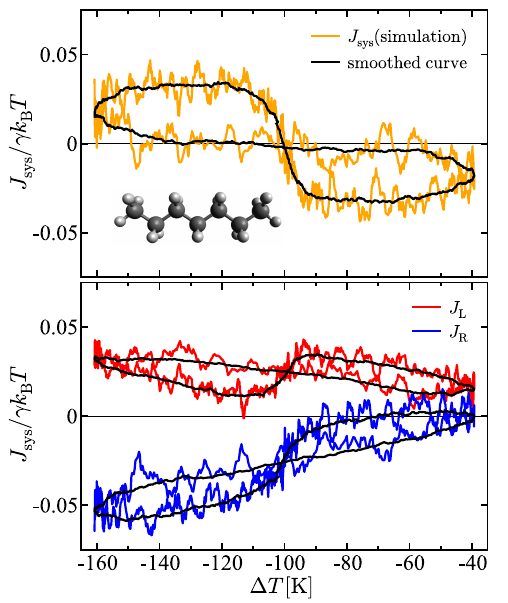}
\caption{\label{fig:septane_Hys_2}
System energy flux (top) and the energy fluxes of the baths (bottom) as a function of temperature difference for an ethane molecular junction.
The parameters in the model are $N =7$, $\gamma_\text{L} \approx 0.05$, $\gamma_\text{R} \approx 0.05$, $T^{(0)}_\text{L} = 200$, $T^{(0)}_\text{R} = 300$,
$\Delta T_\text{L} = 20$, $\Delta T_\text{R} = 50$, $\omega_\text{L} =10$, $\omega_\text{R} = 5$.
The data was vertically shifted by a factor $\Delta J_\text{sys}/\gamma k_\text{B} T = 0.025$.
}
\end{figure}

Shown in Fig.~\ref{fig:septane_Hys_2} are the energy fluxes for the system (top panel) and the baths (bottom panel)
using the same setup as in Fig.~\ref{fig:ethane_Hys_2} but with a longer $N=7$
alkane bridge. The same general features are observed as in the shorter alkane bridge, specially a hysteresis loop with a single pinch. 
The primary difference in the shorter and longer junction is in the magnitude of the hysteresis curve. 
Specifically for $N=7$ the maximum magnitude of the system energy flux is $\approx 0.03$  while the 
$N=2$ the corresponding  maximum magnitude is $\approx 0.1$.
In this case, the longer junction tends to have less pronounced hysteretic behavior while the shorter junction exhibits prominent effects. 
Changing the length of the bridge affects the magnitude of the hysteresis loops but 
has little effect on the loop geometry.
Single-pinch hysteresis loops are observed for both baths as shown in the bottom panel of Fig.~\ref{fig:septane_Hys_2}, with the shape of these curves being similar to the $N=2$ result.

\begin{figure}[]
\includegraphics[width = 8.6cm,clip]{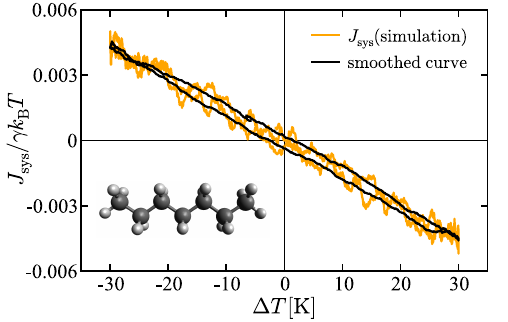}
\caption{\label{fig:septane_Hys_2_quasi}
System energy flux as a function of temperature difference for a septane molecular junction.
The parameters in the model are $N =7$, $\gamma_\text{L} =1$, $\gamma_\text{R} =1$, $T^{(0)}_\text{L} = 300$, $T^{(0)}_\text{R} = 300$,
$\Delta T_\text{L} = 0$, $\Delta T_\text{R} = 30$, $\omega_\text{L} =0$, $\omega_\text{R} = 5 \pi / 6 \approx 2.62$.
The data was vertically shifted by a factor $\Delta J_\text{sys}/\gamma k_\text{B} T = 0.0018$.
}
\end{figure}

The quasi-static (QS) regime of the model is approached when the bath temperatures oscillate slowly with respect to the system-bath couplings  ($\omega_\text{L}/\gamma, \omega_\text{R}/\gamma \rightarrow 0,0)$.
In this regime, hysteresis effects will disappear as the dynamic phase lag between the temperature driving and the system response vanishes. The results of simulations for a model approaching the QS regime are shown in Fig.~\ref{fig:septane_Hys_2_quasi}.
This result was obtained by averaging over $\sim2 \times 10^6$ trajectories.
Here, the system approaches a limit in which hysteresis effects begin to vanish. 
This can be observed by noting that both the amplitude of oscillations as signified through the $y$-axis bounds and the width of hysteresis loop are becoming small.

\subsection{Reduced-Order Model}
We also investigate the molecular junction transport dynamics using a one-dimensional lattice model.
The goal is to develop a reduced-order model 
that captures the important features of the full alkane simulation model.
Reduced-order models and low-dimensional models have played an important historical role in theoretical descriptions of heat transport 
\cite{Isaeva2019modeling,Lepri2003,BonettoFourier2000,Bonetto2004SCR,Segal2009SCR,Lebowitz1959,Lebowitz1967,Lebowitz1971,Segal2021,segal2003thermal,Simine2010FPUT,CampbellFPUT2005,Dhar2008,dhar2006heat}
and we have previously used these types of models to describe nanoscale heat transport hysteresis.  \cite{craven2023b, craven2024a,Chen2024, craven2025entropy} 
Our specific aim here is to model the alkane molecular junction as a harmonic chain and examine 
how and to what extent nonlinearities in the full alkane model manifest in heat transport hysteresis curves.
We will do this using exact analytical expressions for the energy fluxes $J_\text{sys}$, $J_\text{L}$, and $J_\text{R}$
in the harmonic chain and fitting the parameters in those expressions to the simulation data from the full alkane model. 
In essence, a harmonic chain with calibrated parameters is used as a linear surrogate model for the full nonlinear alkane model considered previously.

The specific model considered is a one-dimensional harmonic chain
of $N$ particles. The terminal particles in the chain are each 
connected to heat baths with temperatures that are oscillating in time, in the same fashion 
as the previous MD model.
The Langevin equations of motion for this system are:
\begin{equation}
\begin{aligned}
\label{eq:EoM1har}
m \ddot{x}_1 &= -(k + k_\text{pin}) x_1 + k x_2 - m \gamma_\text{L} \dot{x}_1 + \xi_\text{L}(t), \\
\ldots \\
m \ddot{x}_j  &= -2kx_j+kx_{j-1}+kx_{j+1}, \\
\ldots \\
m \ddot{x}_N &= -(k + k_\text{pin}) x_N + k x_{N-1} - m \gamma_\text{R} \dot{x}_N + \xi_\text{R}(t),
\end{aligned}
\end{equation}
where $m$ is the particle mass (taken to be the mass of a carbon atom), 
$x_j$ is the displacement of particle $j$ from its equilibrium position,
and $k$ is a force constant between particles.
The force constant $k$ can be fit to MD simulation data.
A pinning potential is applied to the two terminal carbons, and $k_\text{pin}$ is the force constant for that potential.  
The parameters $\gamma_\text{L}$ and $\gamma_\text{L}$ are coupling constants between the system and the respective baths and 
$\xi_\text{L}(t)$ and $\xi_\text{R}(t)$ are stochastic forces that obey the scalar versions of the correlations in Eq.~(\ref{eq:FD_theorem_t}).

It is useful to work in frequency space  as we are only concerned with the TDNS of Eq.~(\ref{eq:EoM1har}) and not in any transient behaviors that occur as the system relaxes to this state.
We follow the derivation presented in Ref.~\citenum{craven2024a}, with parts of that derivation repeated here to provide a complete mathematical exposition. Using matrix notation, the equations of motion for the particles in the harmonic molecular chain can be written as
\begin{align}
\label{eq:Langevin_chain_FT_M}
\mathbf{\tilde X}(\omega)&=\mathbf{G}(\omega)\mathbf{\Xi}(\omega),\\
\mathbf{G}(\omega)&=\left[-\mathbf{M}\omega^2+\mathbf{\Phi} + i \boldsymbol{\Gamma}_\text{L}(\omega) + i \boldsymbol{\Gamma}_\text{R}(\omega)\right]^{-1},
\end{align}
where $\tilde{\mathbf{X}}^T=\{\tilde x_1, \tilde x_2, \cdots, \tilde x_N\}$ is a Fourier-transformed displacement vector obtained using $\mathbf{X}(t)=\int_{-\infty}^{\infty}d\omega\tilde{\mathbf{X}}(\omega)e^{i\omega t}$.
The stochastic forces represented in vector form are $\mathbf{\Xi}_\text{L}^T=\{\tilde\xi_\text{L},0, 0, \cdots \}$ and $\mathbf{\Xi}_\text{R}^T=\{0,0,\cdots,\tilde\xi_\text{R}\}$ and we define
$\mathbf{\Xi}(\omega)=\mathbf{\Xi}_\text{L}(\omega)+\mathbf{\Xi}_\text{R}(\omega)$ for notational simplicity.
The function $\boldsymbol{G}(\omega)$ is a Green's function
with conjugate adjoint $\boldsymbol{G}^\dag(\omega)$. 
The matrix $\textbf{M}$ is a diagonal mass matrix, $\boldsymbol{\Phi}$ is a force constant matrix,
with $\boldsymbol{\Gamma}_\text{L}(\omega)$ and $\boldsymbol{\Gamma}_\text{R}(\omega)$ being spectral functions of the heat baths that connect to the system.
The spectral functions used here are $N \times N$ sparse matrices
with element $\boldsymbol{\Gamma}_\text{L}(\omega)_{00}=m\omega\gamma_\text{L}$ 
and element $\boldsymbol{\Gamma}_\text{R}(\omega)_{NN}=m\omega\gamma_\text{R}$,
while all other elements in the matrices are zero.

In matrix notation, the energy fluxes can be expressed as
\begin{align}
	\label{eq:current_left}
 	J_\text{L/R}(t) &= \left\langle (\mathbf{\Lambda}_\text{L/R}\dot{\mathbf{X}})^T \dot{\mathbf{X}}\right\rangle - \left\langle \dot{\mathbf{X}}^T\mathbf{\Xi}_\text{L/R}\right\rangle, \\
 	J_\text{sys}(t) &= -\big(J_\text{L}(t)+J_\text{R}(t)\big).
\end{align}
with $\mathbf{\Lambda}_\text{L}=\mathbf{\Gamma}_\text{L}/\omega$ and $\mathbf{\Lambda}_\text{R}=\mathbf{\Gamma}_\text{R}/\omega$
where the expression for $J_\text{sys}(t)$ arises from conservation of energy.
Evaluating the the correlation functions on the RHS of Eq.~(\ref{eq:current_left}) leads to: \cite{craven2024a}
\begin{widetext}
\begin{align}
\label{eq:left_flux}
	\nonumber J_\text{L/R}(t) 
	=&\int_{-\infty}^{\infty}\omega^2 \frac{k_\text{B}T^{(0)}_\text{L/R}}{\pi}d\omega\text{Tr}[\mathbf{G}^\dagger(\omega)\mathbf{\Lambda}_\text{L/R}^T\mathbf{G}(\omega)\mathbf{\Lambda}_\text{L/R}] +\int_{-\infty}^{\infty}\omega^2 \frac{k_\text{B}T^{(0)}_\text{R/L}}{\pi}d\omega\text{Tr}[\mathbf{G}^\dagger(\omega)\mathbf{\Lambda}_\text{L/R}^T\mathbf{G}(\omega)\mathbf{\Lambda}_\text{R/L}]  \\  \nonumber
	&-\int_{-\infty}^{\infty}\omega (\omega+\omega_\text{L/R})e^{-i\omega_\text{L/R} t}\frac{k_\text{B}\Delta T_\text{L/R}}{2\pi i}d\omega\text{Tr}[\mathbf{G}^\dagger(\omega+\omega_\text{L/R})\mathbf{\Lambda}_\text{L/R}^T\mathbf{G}(\omega)\mathbf{\Lambda}_\text{L/R}]  \\ \nonumber
 	&-\int_{-\infty}^{\infty}\omega (\omega+\omega_\text{R/L})e^{-i\omega_\text{R/L} t}\frac{k_\text{B}\Delta T_\text{R/L}}{2\pi i}d\omega\text{Tr}[\mathbf{G}^\dagger(\omega+\omega_\text{R/L})\mathbf{\Lambda}_\text{L/R}^T\mathbf{G}(\omega)\mathbf{\Lambda}_\text{R/L}]  \\ \nonumber
	&+\int_{-\infty}^{\infty}\omega (\omega-\omega_\text{L/R})e^{i\omega_\text{L/R} t}\frac{k_\text{B}\Delta T_\text{L/R}}{2\pi i}d\omega\text{Tr}[\mathbf{G}^\dagger(\omega-\omega_\text{L/R})\mathbf{\Lambda}_\text{L/R}^T\mathbf{G}(\omega)\mathbf{\Lambda}_\text{L/R}]  \\ \nonumber
	&+\int_{-\infty}^{\infty}\omega (\omega-\omega_\text{R/L})e^{i\omega_\text{R/L} t}\frac{k_\text{B}\Delta T_\text{R/L}}{2\pi i}d\omega\text{Tr}[\mathbf{G}^\dagger(\omega-\omega_\text{R/L})\mathbf{\Lambda}_\text{L/R}^T\mathbf{G}(\omega)\mathbf{\Lambda}_\text{R/L}]  \\ 
	&+i\frac{k_\text{B} T_\text{L/R}(t)}{\pi}\int_{-\infty}^{\infty}\omega d\omega\text{Tr}[\mathbf{\Lambda}_\text{L/R}\mathbf{G}^\dagger(\omega)], \\[1ex]
\label{eq:system_flux}
	\nonumber J_\text{sys}(t)
	=&-\int_{-\infty}^{\infty}\omega^2 \frac{k_\text{B}T^{(0)}_\text{L}}{\pi}d\omega\text{Tr}[\mathbf{G}^\dagger(\omega)\mathbf{\Lambda}^T\mathbf{G}(\omega)\mathbf{\Lambda}_\text{L}] -\int_{-\infty}^{\infty}\omega^2 \frac{k_\text{B}T^{(0)}_\text{R}}{\pi}d\omega\text{Tr}[\mathbf{G}^\dagger(\omega)\mathbf{\Lambda}^T\mathbf{G}(\omega)\mathbf{\Lambda}_\text{R}] \\ \nonumber
	&+\int_{-\infty}^{\infty}\omega (\omega+\omega_\text{L})e^{-i\omega_\text{L} t}\frac{k_\text{B}\Delta T_\text{L}}{2\pi i}d\omega\text{Tr}[\mathbf{G}^\dagger(\omega+\omega_\text{L})\mathbf{\Lambda}^T\mathbf{G}(\omega)\mathbf{\Lambda}_\text{L}]  \\ \nonumber
	&-\int_{-\infty}^{\infty}\omega (\omega-\omega_\text{L})e^{i\omega_\text{L} t}\frac{k_\text{B}\Delta T_\text{L}}{2\pi i}d\omega\text{Tr}[\mathbf{G}^\dagger(\omega-\omega_\text{L})\mathbf{\Lambda}^T\mathbf{G}(\omega)\mathbf{\Lambda}_\text{L}]  \\ \nonumber
	&+\int_{-\infty}^{\infty}\omega (\omega+\omega_\text{R})e^{-i\omega_\text{R} t}\frac{k_\text{B}\Delta T_\text{R}}{2\pi i}d\omega\text{Tr}[\mathbf{G}^\dagger(\omega+\omega_\text{R})\mathbf{\Lambda}^T\mathbf{G}(\omega)\mathbf{\Lambda}_\text{R}]  \\ \nonumber
	&-\int_{-\infty}^{\infty}\omega (\omega-\omega_\text{R})e^{i\omega_\text{R} t}\frac{k_\text{B}\Delta T_\text{R}}{2\pi i}d\omega\text{Tr}[\mathbf{G}^\dagger(\omega-\omega_\text{R})\mathbf{\Lambda}^T\mathbf{G}(\omega)\mathbf{\Lambda}_\text{R}]  \\ 
	&-i\frac{k_\text{B}}{\pi} \left(T_\text{L}(t)\int_{-\infty}^{\infty}\omega d\omega\text{Tr}[\mathbf{\Lambda}_\text{L}\mathbf{G}^\dagger(\omega)]
	+T_\text{R}(t)\int_{-\infty}^{\infty}\omega d\omega \text{Tr}[\mathbf{\Lambda}_\text{R}\mathbf{G}^\dagger(\omega)]\right),
\end{align}
\end{widetext}
where $\mathbf{\Lambda}=\mathbf{\Lambda}_\text{L}+\mathbf{\Lambda}_\text{R}$.

To model an alkane junction, the force constant $k$ in the reduced-order model can be calibrated to the MD simulation 
results for a specific length alkane, i.e., for a specific $N$. We assume that the force constant is independent of the bath parameters. After calibration, the reduced model can be used to quickly and efficiently explore the design space of an alkane junction driven by temperature modulations. The specific calibration procedure employed is to solve the minimization problem
\begin{equation}
k = \underset{k}{\mathrm{argmin}} (\mathcal{L}_\text{sys}(k)),
\end{equation}
where $\mathcal{L}_\text{sys}$ is a loss function defined by the root mean square error (RMSE) between the system energy flux calculated in MD simulations and the system energy flux calculated using the reduced-order model. 
When calibrating to the MD simulations, all other parameters besides $k$ are taken to be the same in both models.
The force constant optimization problem is solved using a grid search approach coupled with a bisection method.

\begin{figure}[]
\includegraphics[width = 8.6cm,clip]{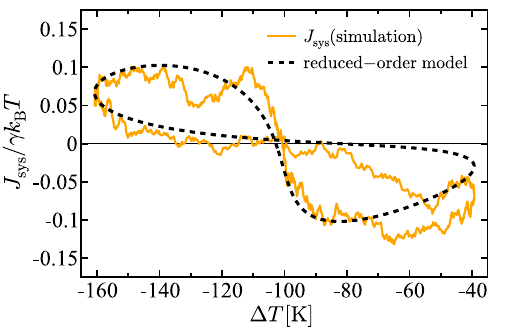}
\caption{\label{fig:RO_Hys_1}
Comparison between the system energy flux calculated using MD simulations (orange) and the corresponding result generated using a calibrated reduced-order model (dashed black). 
The parameters in the model are $N =2$, $\gamma_\text{L} \approx 0.05$, $\gamma_\text{R} \approx 0.05$, $T^{(0)}_\text{L} = 200$, $T^{(0)}_\text{R} = 300$,
$\Delta T_\text{L} = 20$, $\Delta T_\text{R} = 50$, $\omega_\text{L} =10$, $\omega_\text{R} = 5$.
}
\end{figure}

An example result of the calibration procedure is illustrated  Fig.~\ref{fig:RO_Hys_1}. 
The orange noisy data are the results for an ethane molecular junction and the dashed black curve is
the calibrated result from the reduced-order model.
Good agreement is observed between the two results, illustrating the 
the ability of linear low-dimensional models to capture the salient features of the high-dimensional nonlinear simulations.
This is an example case in which the reduced model can well approximate the full model, but we have found in many situations,
for example for the model shown in Fig.~\ref{fig:ethane_Hys_1},
that the nonlinear results cannot be well approximated using the linear models due to appearance of multi-pinch hysteresis loop patterns and other geometric features.

\begin{figure}[]
\includegraphics[width = 8.6cm,clip]{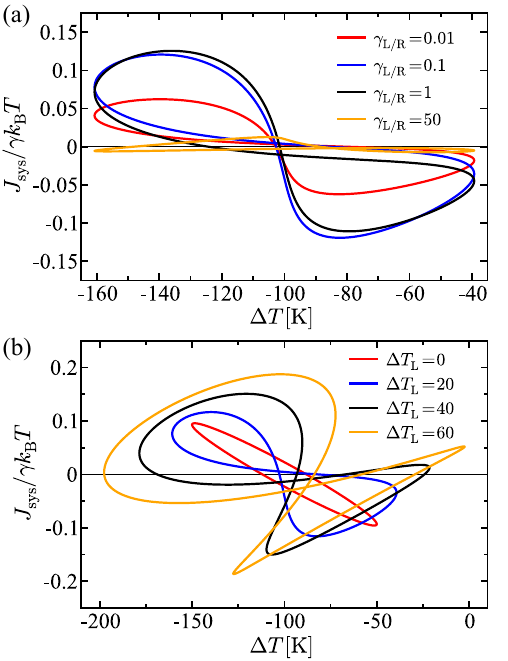}
\caption{\label{fig:RO_gamma}
System energy flux as a function of temperature difference for (a) varying system-bath coupling values and (b) varying temperature oscillation amplitudes $\Delta T_\text{L}$ of the left bath.  
The parameters in the model are $N =2$,  $T^{(0)}_\text{L} = 200$, $T^{(0)}_\text{R} = 300$,
$\Delta T_\text{R} = 50$, $\omega_\text{L} =10$, $\omega_\text{R} = 5$.
In panel (a) $\Delta T_\text{L} = 20$ and in panel (b) $\gamma_\text{L} = \gamma_\text{R} \approx 0.05$.
}
\end{figure}

Once the force constant is calibrated to the MD results, the model parameter space can be efficiently explored. 
First, we examine how changing the system bath couplings of the two baths
$\gamma_\text{L/R}$ affects hysteresis.
Shown in Fig.~\ref{fig:RO_gamma}(a) are the results of the calibrated reduced-order model.
Increasing $\gamma_\text{L/R}$ generally results in larger magnitudes of oscillation in the hysteresis loop. 
This is because energy transfer between the baths is supported by the increased coupling strength,
and so increasing the coupling leads to more pronounced energy transfer dynamics. 
However, it should be noted that the size of the hysteresis loops is not a monotonically increasing function of $\gamma_\text{L/R}$, and that, instead, there is a distinctive turnover behavior that occurs. This can be seen in the  $\gamma_\text{L} = 50$ curve which is smaller than the other loops in the scaled units shown. As 
$\gamma_\text{L/R}$ are increased well beyond the temperature oscillation frequency, the quasistatic limit is approached and the width of the hysteresis loops goes to zero.

Figure~\ref{fig:RO_gamma}(b) shows the result of varying $\Delta T_\text{L}$ while keep all other parameters constant. 
Interestingly, increasing $\Delta T_\text{L}$  does not affect the hysteresis shape or magnitude in a linear manner but instead leads to complex and nonlinear effects. For $\Delta T_\text{L} = 0$ only one bath is oscillating and a simple oval hysteresis pattern is observed. When the driving amplitude is increased to $\Delta T_\text{L} = 20$, we recover the same results shown in Fig.~\ref{fig:ethane_Hys_2}.
For $\Delta T_\text{L} = 40$, a highly asymmetrical loop pattern emerges, and the same general asymmetrical hysteresis behavior is observed but in an exacerbated manner for $\Delta T_\text{L} = 60$.
As $\Delta T_\text{L}$ is increased, the magnitude of the hysteresis curves generally increases. 
For example, the maximum amplitude for $\Delta T_\text{L} = 0 $ is $\approx 0.1$  while the maximum amplitude for $\Delta T_\text{L} = 60$ is $\approx 0.2$.
The sign change patterns for $J_\text{sys}$ are increasingly complex for increasing $\Delta T_\text{L}$, and for larger driving amplitudes the sign of $J_\text{sys}$ can change in complex manner throughout the oscillation cycle. 
Overall, the complex behavior of the model under varying $\Delta T_\text{L}$ highlights the complicated dynamics that can emerge in thermally-driven molecular junctions.

\begin{figure}[]
\includegraphics[width = 8.6cm,clip]{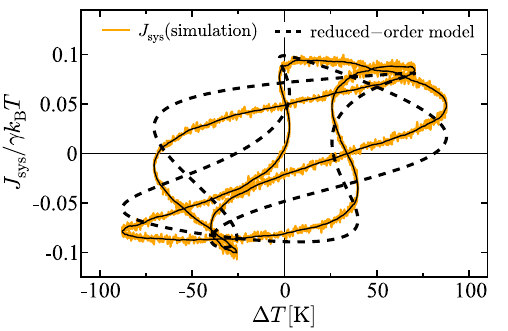}
\caption{\label{fig:Hys_1_Methane}
Comparison between the system energy flux calculated using MD simulations (orange) and the corresponding result generated using a calibrated reduced-order model (dashed black). The solid black curve is the simulation result with smoothing to guide the eye.
The parameters in the model are $N =1$, $\gamma_\text{L} = 1.5$, $\gamma_\text{R} =0.2$, $T^{(0)}_\text{L} = 300$, $T^{(0)}_\text{R} = 300$,
$\Delta T_\text{L} = 60$, $\Delta T_\text{R} = 30$, $\omega_\text{L} =2$, $\omega_\text{R} = 5$.
}
\end{figure}

Figure~\ref{fig:Hys_1_Methane} shows the results of a methane junction for the case 
in which the temperatures of both baths are oscillating, but at different frequencies and with asymmetric system-bath couplings. 
The hysteresis pattern generated is complex, showing an interweaving path and an asymmetric shape. 
In this complex system, the reduced-order model (shown as a dashed black curve) again captures the qualitative behavior of the highly complex hysteresis curve.
This result suggests that reduced-order models can be used to understand, qualitatively, time-dependent heat transport behavior in alkane molecular junctions in certain physical regimes.

\begin{figure}[]
\includegraphics[width = 8.6cm,clip]{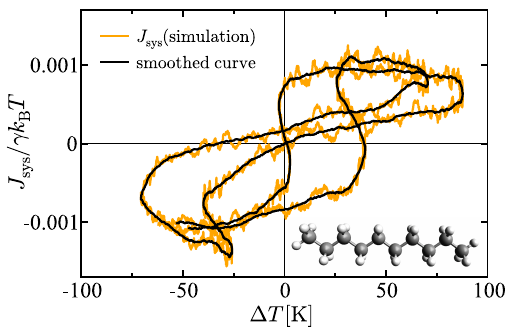}
\caption{\label{fig:Hys_1_Decane}
System energy flux calculated using MD simulations (orange) for a decane molecular junction. The solid black curve is the simulation result with smoothing.
The parameters in the model are $N =10$, $\gamma_\text{L} = 1.5$, $\gamma_\text{R} =0.2$, $T^{(0)}_\text{L} = 300$, $T^{(0)}_\text{R} = 300$,
$\Delta T_\text{L} = 60$, $\Delta T_\text{R} = 30$, $\omega_\text{L} =2$, $\omega_\text{R} = 5$. The data was vertically shifted by a factor $\Delta J_\text{sys}/\gamma k_\text{B} T = 0.000625$.}
\end{figure}

Figure~\ref{fig:Hys_1_Decane} illustrates the results for a decane ($N=10$) junction using the same system parameters (multi-frequency driving) as
in Fig.~\ref{fig:Hys_1_Methane}.
Again, the hysteresis curve is highly complex showing multiple overlap points and an interwoven pattern.
Comparing the results in Fig.~\ref{fig:Hys_1_Decane} to those in Fig.~\ref{fig:Hys_1_Methane}, we see that the heat transport magnitude is significantly decreased in the longer junction, but that the hysteresis curves are qualitatively similar in shape. This implies two principal things: (1) shorter junctions may be a better choice experimentally to elucidate heat transport hysteresis because the magnitude of the heat transport is larger and (2) the length of the junction does affect heat transport, primarily in the magnitude, but generally qualitative behaviors in the hysteresis patterns persist across length scales. Further work to characterize the transition from  ballistic to diffusive transport and how hysteresis manifests in those regimes is currently underway.

\section{Conclusions}
\label{sec:conclusion}

This work has demonstrated using theoretical models that molecular junctions can exhibit heat transport hysteresis when subjected to a time-periodic temperature gradient. 
This phenomenon reveals a distinct memory effect in energy transport processes at the nanoscale that can potentially be utilized for 
advanced energy storage devices and neuromorphic computing applications.
By employing nonequilibrium molecular dynamics simulations and stochastic thermodynamics, we have examined this effect across a range of system parameters, including over different temperature oscillation frequencies and chain lengths. The presented findings illustrate the potential for molecular junctions--a well-studied nanoscale system---to serve as components in nanoscale thermal neuromorphic computing architectures. 
In analogy to memristive or memcapacitive devices in electronic neuromorphic systems, molecular junctions that exhibit heat transport hysteresis can store and process information through complex power cycling loops and hysteresis loops. This effect is absent 
under steady-state conditions and only occurs when the system is driven into a time-dependent nonequilibrium state by an oscillating temperature gradient. By leveraging this history-dependent response, it may be possible to design thermal circuits and logic elements that perform both memory and processing tasks, i.e., thermal neuromorphic computers.
Molecular junctions with tunable hysteresis characteristics could, in principle, serve as thermal analogs to electronic memristors and memcapacitors, enabling nanoscale thermal memory and logic operations driven by oscillating temperature gradients in a similar fashion to how electronic memristive behavior is driven by an oscillating voltage.

There are several key takeaways from this work:
\begin{enumerate}

\item Nonlinearities can give rise to complex heat transport hysteresis loop patterns that include effects that cannot be generated using linear models.
\item However, hysteresis is not purely a nonlinear effect, and can be observed in linear (harmonic) models and systems. Moreover, in some cases, linear models can be used to well-approximate the hysteresis dynamics of nonlinear systems.
\item Analytical results will play an important role in the analysis of heat transport hysteresis 
as it requires significant computational resources to converge the energy flux calculations and to differentiate the energy flux signals from thermal noise in nanoscale systems.
\item The length of a molecular junction affects the geometry of the heat transport hysteresis loops. 
\item The observed thermal memory behaviors exist in a physical regime not fully captured by the conventional memristive definition in which a pinch point in the  hysteresis loop occurs at $\Delta T = 0$.
\end{enumerate}

Attempting to observe heat transport hysteresis in single-molecule junction experiments is an important next step. 
From a computational perspective, developing nonlinear surrogate models will enable fast and systematic probing of the model parameter space.
This could, in turn, lead to a more complete understanding of the physical mechanisms driving hysteresis.
In this work, we have used an alkane molecular junction as a representative system. However, altering the molecular structure of a junction through, for example,  chemical functionalization, can alter heat transport. Exploring other molecule types besides alkanes will therefore be important in future work.
Another important direction for future work is to explore quantum effects in heat transport hysteresis, such as those studied in recent work on heat pulses in electron quantum optics.\cite{Portugal2024prl} Extending the present framework to incorporate quantum statistics or coherent transport dynamics could reveal new hysteresis regimes and phenomena relevant to thermotronics and thermal computing.

\section{Acknowledgments}
This article is contributed in honor of Abe Nitzan.
We acknowledge support from the Los Alamos National Laboratory (LANL) 
Directed Research and Development funds (LDRD).
The computing resources used to perform this
research were provided by the LANL Institutional Computing Program.

\bibliographystyle{apsrev}

\begin{thebibliography}{107}
\expandafter\ifx\csname natexlab\endcsname\relax\def\natexlab#1{#1}\fi
\expandafter\ifx\csname bibnamefont\endcsname\relax
  \def\bibnamefont#1{#1}\fi
\expandafter\ifx\csname bibfnamefont\endcsname\relax
  \def\bibfnamefont#1{#1}\fi
\expandafter\ifx\csname citenamefont\endcsname\relax
  \def\citenamefont#1{#1}\fi
\expandafter\ifx\csname url\endcsname\relax
  \def\url#1{\texttt{#1}}\fi
\expandafter\ifx\csname urlprefix\endcsname\relax\def\urlprefix{URL }\fi
\providecommand{\bibinfo}[2]{#2}
\providecommand{\eprint}[2][]{\url{#2}}

\bibitem[{\citenamefont{Cahill et~al.}(2002)\citenamefont{Cahill, Goodson, and Majumdar}}]{Cahill2002}
\bibinfo{author}{\bibfnamefont{D.~G.} \bibnamefont{Cahill}}, \bibinfo{author}{\bibfnamefont{K.}~\bibnamefont{Goodson}}, \bibnamefont{and} \bibinfo{author}{\bibfnamefont{A.}~\bibnamefont{Majumdar}}, \bibinfo{journal}{J. Heat Transfer} \textbf{\bibinfo{volume}{124}}, \bibinfo{pages}{223} (\bibinfo{year}{2002}).

\bibitem[{\citenamefont{Cahill et~al.}(2003)\citenamefont{Cahill, Ford, Goodson, Mahan, Majumdar, Maris, Merlin, and Phillpot}}]{Cahill2003}
\bibinfo{author}{\bibfnamefont{D.~G.} \bibnamefont{Cahill}}, \bibinfo{author}{\bibfnamefont{W.~K.} \bibnamefont{Ford}}, \bibinfo{author}{\bibfnamefont{K.~E.} \bibnamefont{Goodson}}, \bibinfo{author}{\bibfnamefont{G.~D.} \bibnamefont{Mahan}}, \bibinfo{author}{\bibfnamefont{A.}~\bibnamefont{Majumdar}}, \bibinfo{author}{\bibfnamefont{H.~J.} \bibnamefont{Maris}}, \bibinfo{author}{\bibfnamefont{R.}~\bibnamefont{Merlin}}, \bibnamefont{and} \bibinfo{author}{\bibfnamefont{S.~R.} \bibnamefont{Phillpot}}, \bibinfo{journal}{J. Appl. Phys.} \textbf{\bibinfo{volume}{93}}, \bibinfo{pages}{793} (\bibinfo{year}{2003}), \eprint{doi:10.1063/1.1524305}.

\bibitem[{\citenamefont{Dhar}(2008)}]{Dhar2008}
\bibinfo{author}{\bibfnamefont{A.}~\bibnamefont{Dhar}}, \bibinfo{journal}{Adv. Phys.} \textbf{\bibinfo{volume}{57}}, \bibinfo{pages}{457} (\bibinfo{year}{2008}), \eprint{doi:10.1080/00018730802538522}.

\bibitem[{\citenamefont{Dubi and Di~Ventra}(2011)}]{Dubi2011}
\bibinfo{author}{\bibfnamefont{Y.}~\bibnamefont{Dubi}} \bibnamefont{and} \bibinfo{author}{\bibfnamefont{M.}~\bibnamefont{Di~Ventra}}, \bibinfo{journal}{Rev. Mod. Phys.} \textbf{\bibinfo{volume}{83}}, \bibinfo{pages}{131} (\bibinfo{year}{2011}), \eprint{doi:10.1103/RevModPhys.83.131}.

\bibitem[{\citenamefont{Narayana and Sato}(2012)}]{Sato2012}
\bibinfo{author}{\bibfnamefont{S.}~\bibnamefont{Narayana}} \bibnamefont{and} \bibinfo{author}{\bibfnamefont{Y.}~\bibnamefont{Sato}}, \bibinfo{journal}{Phys. Rev. Lett.} \textbf{\bibinfo{volume}{108}}, \bibinfo{pages}{214303} (\bibinfo{year}{2012}), \eprint{doi:10.1103/PhysRevLett.108.214303}.

\bibitem[{\citenamefont{Maldovan}(2013)}]{Maldovan2013}
\bibinfo{author}{\bibfnamefont{M.}~\bibnamefont{Maldovan}}, \bibinfo{journal}{Nature} \textbf{\bibinfo{volume}{503}}, \bibinfo{pages}{209} (\bibinfo{year}{2013}), \eprint{doi:10.1038/nature12608}.

\bibitem[{\citenamefont{Segal and Agarwalla}(2016)}]{Segal2016}
\bibinfo{author}{\bibfnamefont{D.}~\bibnamefont{Segal}} \bibnamefont{and} \bibinfo{author}{\bibfnamefont{B.~K.} \bibnamefont{Agarwalla}}, \bibinfo{journal}{Annu. Rev. Phys. Chem.} \textbf{\bibinfo{volume}{67}}, \bibinfo{pages}{185} (\bibinfo{year}{2016}), \eprint{doi:10.1146/annurev-physchem-040215-112103}.

\bibitem[{\citenamefont{Ness et~al.}(2016)\citenamefont{Ness, Genina, Stella, Lorenz, and Kantorovich}}]{Ness2016}
\bibinfo{author}{\bibfnamefont{H.}~\bibnamefont{Ness}}, \bibinfo{author}{\bibfnamefont{A.}~\bibnamefont{Genina}}, \bibinfo{author}{\bibfnamefont{L.}~\bibnamefont{Stella}}, \bibinfo{author}{\bibfnamefont{C.~D.} \bibnamefont{Lorenz}}, \bibnamefont{and} \bibinfo{author}{\bibfnamefont{L.}~\bibnamefont{Kantorovich}}, \bibinfo{journal}{Phys. Rev. B} \textbf{\bibinfo{volume}{93}}, \bibinfo{pages}{174303} (\bibinfo{year}{2016}), \eprint{doi:10.1103/PhysRevB.93.174303}.

\bibitem[{\citenamefont{Ness et~al.}(2017)\citenamefont{Ness, Stella, Lorenz, and Kantorovich}}]{Ness2017}
\bibinfo{author}{\bibfnamefont{H.}~\bibnamefont{Ness}}, \bibinfo{author}{\bibfnamefont{L.}~\bibnamefont{Stella}}, \bibinfo{author}{\bibfnamefont{C.~D.} \bibnamefont{Lorenz}}, \bibnamefont{and} \bibinfo{author}{\bibfnamefont{L.}~\bibnamefont{Kantorovich}}, \bibinfo{journal}{J. Chem. Phys.} \textbf{\bibinfo{volume}{146}} (\bibinfo{year}{2017}), \eprint{doi:10.1063/1.4981816}.

\bibitem[{\citenamefont{Nascimento and Morgado}(2022)}]{Nascimento2022}
\bibinfo{author}{\bibfnamefont{E.~S.} \bibnamefont{Nascimento}} \bibnamefont{and} \bibinfo{author}{\bibfnamefont{W.~A.~M.} \bibnamefont{Morgado}}, \bibinfo{journal}{J. Phys. A Math. Theor.} \textbf{\bibinfo{volume}{55}}, \bibinfo{pages}{395003} (\bibinfo{year}{2022}), \eprint{doi:10.1088/1751-8121/ac8c07}.

\bibitem[{\citenamefont{Li et~al.}(2012)\citenamefont{Li, Ren, Wang, Zhang, H\"anggi, and Li}}]{Li2012}
\bibinfo{author}{\bibfnamefont{N.}~\bibnamefont{Li}}, \bibinfo{author}{\bibfnamefont{J.}~\bibnamefont{Ren}}, \bibinfo{author}{\bibfnamefont{L.}~\bibnamefont{Wang}}, \bibinfo{author}{\bibfnamefont{G.}~\bibnamefont{Zhang}}, \bibinfo{author}{\bibfnamefont{P.}~\bibnamefont{H\"anggi}}, \bibnamefont{and} \bibinfo{author}{\bibfnamefont{B.}~\bibnamefont{Li}}, \bibinfo{journal}{Rev. Mod. Phys.} \textbf{\bibinfo{volume}{84}}, \bibinfo{pages}{1045} (\bibinfo{year}{2012}), \eprint{doi:10.1103/RevModPhys.84.1045}.

\bibitem[{\citenamefont{Sabhapandit}(2012{\natexlab{a}})}]{Sabhapandit2012}
\bibinfo{author}{\bibfnamefont{S.}~\bibnamefont{Sabhapandit}}, \bibinfo{journal}{Phys. Rev. E} \textbf{\bibinfo{volume}{85}}, \bibinfo{pages}{021108} (\bibinfo{year}{2012}{\natexlab{a}}), \eprint{doi:10.1103/PhysRevE.85.021108}.

\bibitem[{\citenamefont{Lebowitz}(1959)}]{Lebowitz1959}
\bibinfo{author}{\bibfnamefont{J.~L.} \bibnamefont{Lebowitz}}, \bibinfo{journal}{Phys. Rev.} \textbf{\bibinfo{volume}{114}}, \bibinfo{pages}{1192} (\bibinfo{year}{1959}), \eprint{doi:10.1103/PhysRev.114.1192}.

\bibitem[{\citenamefont{Rieder et~al.}(1967)\citenamefont{Rieder, Lebowitz, and Lieb}}]{Lebowitz1967}
\bibinfo{author}{\bibfnamefont{Z.}~\bibnamefont{Rieder}}, \bibinfo{author}{\bibfnamefont{J.~L.} \bibnamefont{Lebowitz}}, \bibnamefont{and} \bibinfo{author}{\bibfnamefont{E.}~\bibnamefont{Lieb}}, \bibinfo{journal}{J. Math. Phys.} \textbf{\bibinfo{volume}{8}}, \bibinfo{pages}{1073} (\bibinfo{year}{1967}), \eprint{doi:10.1063/1.1705319}.

\bibitem[{\citenamefont{Casher and Lebowitz}(1971)}]{Lebowitz1971}
\bibinfo{author}{\bibfnamefont{A.}~\bibnamefont{Casher}} \bibnamefont{and} \bibinfo{author}{\bibfnamefont{J.~L.} \bibnamefont{Lebowitz}}, \bibinfo{journal}{J. Math. Phys.} \textbf{\bibinfo{volume}{12}}, \bibinfo{pages}{1701} (\bibinfo{year}{1971}), \eprint{doi:10.1063/1.1665794}.

\bibitem[{\citenamefont{Dhar and Lebowitz}(2008)}]{Lebowitz2008}
\bibinfo{author}{\bibfnamefont{A.}~\bibnamefont{Dhar}} \bibnamefont{and} \bibinfo{author}{\bibfnamefont{J.~L.} \bibnamefont{Lebowitz}}, \bibinfo{journal}{Phys. Rev. Lett.} \textbf{\bibinfo{volume}{100}}, \bibinfo{pages}{134301} (\bibinfo{year}{2008}), \eprint{doi:10.1103/PhysRevLett.100.134301}.

\bibitem[{\citenamefont{Kannan et~al.}(2012)\citenamefont{Kannan, Dhar, and Lebowitz}}]{Lebowitz2012}
\bibinfo{author}{\bibfnamefont{V.}~\bibnamefont{Kannan}}, \bibinfo{author}{\bibfnamefont{A.}~\bibnamefont{Dhar}}, \bibnamefont{and} \bibinfo{author}{\bibfnamefont{J.~L.} \bibnamefont{Lebowitz}}, \bibinfo{journal}{Phys. Rev. E} \textbf{\bibinfo{volume}{85}}, \bibinfo{pages}{041118} (\bibinfo{year}{2012}), \eprint{doi:10.1103/PhysRevE.85.041118}.

\bibitem[{\citenamefont{Segal et~al.}(2003{\natexlab{a}})\citenamefont{Segal, Nitzan, and H\"anggi}}]{Nitzan2003thermal}
\bibinfo{author}{\bibfnamefont{D.}~\bibnamefont{Segal}}, \bibinfo{author}{\bibfnamefont{A.}~\bibnamefont{Nitzan}}, \bibnamefont{and} \bibinfo{author}{\bibfnamefont{P.}~\bibnamefont{H\"anggi}}, \bibinfo{journal}{J. Chem. Phys.} \textbf{\bibinfo{volume}{119}}, \bibinfo{pages}{6840} (\bibinfo{year}{2003}{\natexlab{a}}), \eprint{doi:10.1063/1.1603211}.

\bibitem[{\citenamefont{Segal and Nitzan}(2005)}]{Segal2005prl}
\bibinfo{author}{\bibfnamefont{D.}~\bibnamefont{Segal}} \bibnamefont{and} \bibinfo{author}{\bibfnamefont{A.}~\bibnamefont{Nitzan}}, \bibinfo{journal}{Phys. Rev. Lett.} \textbf{\bibinfo{volume}{94}}, \bibinfo{pages}{034301} (\bibinfo{year}{2005}), \eprint{doi:10.1103/PhysRevLett.94.034301}.

\bibitem[{\citenamefont{Dhar and Dandekar}(2015)}]{Dhar2015}
\bibinfo{author}{\bibfnamefont{A.}~\bibnamefont{Dhar}} \bibnamefont{and} \bibinfo{author}{\bibfnamefont{R.}~\bibnamefont{Dandekar}}, \bibinfo{journal}{Physica A} \textbf{\bibinfo{volume}{418}}, \bibinfo{pages}{49 } (\bibinfo{year}{2015}), \eprint{doi:10.1016/j.physa.2014.06.002}.

\bibitem[{\citenamefont{Velizhanin et~al.}(2015)\citenamefont{Velizhanin, Sahu, Chien, Dubi, and Zwolak}}]{Velizhanin2015}
\bibinfo{author}{\bibfnamefont{K.~A.} \bibnamefont{Velizhanin}}, \bibinfo{author}{\bibfnamefont{S.}~\bibnamefont{Sahu}}, \bibinfo{author}{\bibfnamefont{C.-C.} \bibnamefont{Chien}}, \bibinfo{author}{\bibfnamefont{Y.}~\bibnamefont{Dubi}}, \bibnamefont{and} \bibinfo{author}{\bibfnamefont{M.}~\bibnamefont{Zwolak}}, \bibinfo{journal}{Sci. Rep.} \textbf{\bibinfo{volume}{5}} (\bibinfo{year}{2015}), \eprint{doi:10.1038/srep17506}.

\bibitem[{\citenamefont{Murashita and Esposito}(2016)}]{Esposito2016}
\bibinfo{author}{\bibfnamefont{Y.}~\bibnamefont{Murashita}} \bibnamefont{and} \bibinfo{author}{\bibfnamefont{M.}~\bibnamefont{Esposito}}, \bibinfo{journal}{Phys. Rev. E} \textbf{\bibinfo{volume}{94}}, \bibinfo{pages}{062148} (\bibinfo{year}{2016}), \eprint{doi:10.1103/PhysRevE.94.062148}.

\bibitem[{\citenamefont{Craven and Nitzan}(2016)}]{craven16c}
\bibinfo{author}{\bibfnamefont{G.~T.} \bibnamefont{Craven}} \bibnamefont{and} \bibinfo{author}{\bibfnamefont{A.}~\bibnamefont{Nitzan}}, \bibinfo{journal}{Proc. Natl. Acad. Sci.} \textbf{\bibinfo{volume}{113}}, \bibinfo{pages}{9421} (\bibinfo{year}{2016}), \eprint{doi:10.1073/pnas.1609141113}.

\bibitem[{\citenamefont{Matyushov}(2016)}]{matyushov16c}
\bibinfo{author}{\bibfnamefont{D.~V.} \bibnamefont{Matyushov}}, \bibinfo{journal}{Proc. Natl. Acad. Sci.} \textbf{\bibinfo{volume}{113}}, \bibinfo{pages}{9401} (\bibinfo{year}{2016}), \eprint{doi:10.1073/pnas.1610542113}.

\bibitem[{\citenamefont{Craven and Nitzan}(2017{\natexlab{a}})}]{craven17a}
\bibinfo{author}{\bibfnamefont{G.~T.} \bibnamefont{Craven}} \bibnamefont{and} \bibinfo{author}{\bibfnamefont{A.}~\bibnamefont{Nitzan}}, \bibinfo{journal}{J. Chem. Phys.} \textbf{\bibinfo{volume}{146}}, \bibinfo{pages}{092305} (\bibinfo{year}{2017}{\natexlab{a}}), \eprint{doi:10.1063/1.4971293}.

\bibitem[{\citenamefont{Craven and Nitzan}(2017{\natexlab{b}})}]{craven17b}
\bibinfo{author}{\bibfnamefont{G.~T.} \bibnamefont{Craven}} \bibnamefont{and} \bibinfo{author}{\bibfnamefont{A.}~\bibnamefont{Nitzan}}, \bibinfo{journal}{Phys. Rev. Lett.} \textbf{\bibinfo{volume}{118}}, \bibinfo{pages}{207201} (\bibinfo{year}{2017}{\natexlab{b}}), \eprint{doi:10.1103/PhysRevLett.118.207201}.

\bibitem[{\citenamefont{Chen et~al.}(2017)\citenamefont{Chen, Craven, and Nitzan}}]{craven17e}
\bibinfo{author}{\bibfnamefont{R.}~\bibnamefont{Chen}}, \bibinfo{author}{\bibfnamefont{G.~T.} \bibnamefont{Craven}}, \bibnamefont{and} \bibinfo{author}{\bibfnamefont{A.}~\bibnamefont{Nitzan}}, \bibinfo{journal}{J. Chem. Phys.} \textbf{\bibinfo{volume}{147}}, \bibinfo{pages}{124101} (\bibinfo{year}{2017}), \eprint{doi:10.1063/1.4990410}.

\bibitem[{\citenamefont{Craven and Nitzan}(2020)}]{craven20a}
\bibinfo{author}{\bibfnamefont{G.~T.} \bibnamefont{Craven}} \bibnamefont{and} \bibinfo{author}{\bibfnamefont{A.}~\bibnamefont{Nitzan}}, \bibinfo{journal}{Nano Lett.} \textbf{\bibinfo{volume}{20}}, \bibinfo{pages}{989} (\bibinfo{year}{2020}), \eprint{doi:10.1021/acs.nanolett.9b04070}.

\bibitem[{\citenamefont{Ochoa}(2022)}]{Ochoa2022}
\bibinfo{author}{\bibfnamefont{M.~A.} \bibnamefont{Ochoa}}, \bibinfo{journal}{Phys. Rev. E} \textbf{\bibinfo{volume}{106}}, \bibinfo{pages}{064113} (\bibinfo{year}{2022}), \eprint{doi:10.1103/PhysRevE.106.064113}.

\bibitem[{\citenamefont{Bonetto et~al.}(2000)\citenamefont{Bonetto, Lebowitz, and Rey-Bellet}}]{BonettoFourier2000}
\bibinfo{author}{\bibfnamefont{F.}~\bibnamefont{Bonetto}}, \bibinfo{author}{\bibfnamefont{J.~L.} \bibnamefont{Lebowitz}}, \bibnamefont{and} \bibinfo{author}{\bibfnamefont{L.}~\bibnamefont{Rey-Bellet}}, \emph{\bibinfo{title}{Mathematical Physics}} (\bibinfo{publisher}{Imperial College, London}, \bibinfo{year}{2000}), pp. \bibinfo{pages}{128--150}.

\bibitem[{\citenamefont{Bonetto et~al.}(2004)\citenamefont{Bonetto, Lebowitz, and Lukkarinen}}]{Bonetto2004SCR}
\bibinfo{author}{\bibfnamefont{F.}~\bibnamefont{Bonetto}}, \bibinfo{author}{\bibfnamefont{J.~L.} \bibnamefont{Lebowitz}}, \bibnamefont{and} \bibinfo{author}{\bibfnamefont{J.}~\bibnamefont{Lukkarinen}}, \bibinfo{journal}{J. Stat. Phys.} \textbf{\bibinfo{volume}{116}}, \bibinfo{pages}{783} (\bibinfo{year}{2004}), \eprint{doi:10.1023/B:JOSS.0000037232.14365.10}.

\bibitem[{\citenamefont{Segal}(2009)}]{Segal2009SCR}
\bibinfo{author}{\bibfnamefont{D.}~\bibnamefont{Segal}}, \bibinfo{journal}{Phys. Rev. E} \textbf{\bibinfo{volume}{79}}, \bibinfo{pages}{012103} (\bibinfo{year}{2009}), \eprint{doi:10.1103/PhysRevE.79.012103}.

\bibitem[{\citenamefont{Chang et~al.}(2008)\citenamefont{Chang, Okawa, Garcia, Majumdar, and Zettl}}]{Chang2008}
\bibinfo{author}{\bibfnamefont{C.~W.} \bibnamefont{Chang}}, \bibinfo{author}{\bibfnamefont{D.}~\bibnamefont{Okawa}}, \bibinfo{author}{\bibfnamefont{H.}~\bibnamefont{Garcia}}, \bibinfo{author}{\bibfnamefont{A.}~\bibnamefont{Majumdar}}, \bibnamefont{and} \bibinfo{author}{\bibfnamefont{A.}~\bibnamefont{Zettl}}, \bibinfo{journal}{Phys. Rev. Lett.} \textbf{\bibinfo{volume}{101}}, \bibinfo{pages}{075903} (\bibinfo{year}{2008}), \eprint{doi:10.1103/PhysRevLett.101.075903}.

\bibitem[{\citenamefont{Craven and Nitzan}(2023)}]{craven2023a}
\bibinfo{author}{\bibfnamefont{G.~T.} \bibnamefont{Craven}} \bibnamefont{and} \bibinfo{author}{\bibfnamefont{A.}~\bibnamefont{Nitzan}}, \bibinfo{journal}{J. Chem. Phys.} \textbf{\bibinfo{volume}{158}} (\bibinfo{year}{2023}), \eprint{doi:10.1063/5.0144248}.

\bibitem[{\citenamefont{Ratner}(2013)}]{Ratner2013review}
\bibinfo{author}{\bibfnamefont{M.~A.} \bibnamefont{Ratner}}, \bibinfo{journal}{Nat. Nanotechnol.} \textbf{\bibinfo{volume}{8}}, \bibinfo{pages}{378} (\bibinfo{year}{2013}), \eprint{doi:10.1038/nnano.2013.110}.

\bibitem[{\citenamefont{Xiang et~al.}(2016)\citenamefont{Xiang, Wang, Jia, Lee, and Guo}}]{Xiang2016}
\bibinfo{author}{\bibfnamefont{D.}~\bibnamefont{Xiang}}, \bibinfo{author}{\bibfnamefont{X.}~\bibnamefont{Wang}}, \bibinfo{author}{\bibfnamefont{C.}~\bibnamefont{Jia}}, \bibinfo{author}{\bibfnamefont{T.}~\bibnamefont{Lee}}, \bibnamefont{and} \bibinfo{author}{\bibfnamefont{X.}~\bibnamefont{Guo}}, \bibinfo{journal}{Chem. Rev.} \textbf{\bibinfo{volume}{116}}, \bibinfo{pages}{4318} (\bibinfo{year}{2016}), \eprint{doi:10.1021/acs.chemrev.5b00680}.

\bibitem[{\citenamefont{Joulain et~al.}(2016)\citenamefont{Joulain, Drevillon, Ezzahri, and Ordonez-Miranda}}]{Joulain2016}
\bibinfo{author}{\bibfnamefont{K.}~\bibnamefont{Joulain}}, \bibinfo{author}{\bibfnamefont{J.}~\bibnamefont{Drevillon}}, \bibinfo{author}{\bibfnamefont{Y.}~\bibnamefont{Ezzahri}}, \bibnamefont{and} \bibinfo{author}{\bibfnamefont{J.}~\bibnamefont{Ordonez-Miranda}}, \bibinfo{journal}{Phys. Rev. Lett.} \textbf{\bibinfo{volume}{116}}, \bibinfo{pages}{200601} (\bibinfo{year}{2016}), \eprint{doi:10.1103/PhysRevLett.116.200601}.

\bibitem[{\citenamefont{Gehring et~al.}(2021)\citenamefont{Gehring, Sowa, Hsu, de~Bruijckere, van~der Star, Le~Roy, Bogani, Gauger, and van~der Zant}}]{Gehring2021}
\bibinfo{author}{\bibfnamefont{P.}~\bibnamefont{Gehring}}, \bibinfo{author}{\bibfnamefont{J.~K.} \bibnamefont{Sowa}}, \bibinfo{author}{\bibfnamefont{C.}~\bibnamefont{Hsu}}, \bibinfo{author}{\bibfnamefont{J.}~\bibnamefont{de~Bruijckere}}, \bibinfo{author}{\bibfnamefont{M.}~\bibnamefont{van~der Star}}, \bibinfo{author}{\bibfnamefont{J.~J.} \bibnamefont{Le~Roy}}, \bibinfo{author}{\bibfnamefont{L.}~\bibnamefont{Bogani}}, \bibinfo{author}{\bibfnamefont{E.~M.} \bibnamefont{Gauger}}, \bibnamefont{and} \bibinfo{author}{\bibfnamefont{H.~S.} \bibnamefont{van~der Zant}}, \bibinfo{journal}{Nature Nanotechnology} \textbf{\bibinfo{volume}{16}}, \bibinfo{pages}{426} (\bibinfo{year}{2021}).

\bibitem[{\citenamefont{Russ et~al.}(2016)\citenamefont{Russ, Glaudell, Urban, Chabinyc, and Segalman}}]{Russ2016organic}
\bibinfo{author}{\bibfnamefont{B.}~\bibnamefont{Russ}}, \bibinfo{author}{\bibfnamefont{A.}~\bibnamefont{Glaudell}}, \bibinfo{author}{\bibfnamefont{J.~J.} \bibnamefont{Urban}}, \bibinfo{author}{\bibfnamefont{M.~L.} \bibnamefont{Chabinyc}}, \bibnamefont{and} \bibinfo{author}{\bibfnamefont{R.~A.} \bibnamefont{Segalman}}, \bibinfo{journal}{Nat. Rev. Mater.} \textbf{\bibinfo{volume}{1}}, \bibinfo{pages}{1} (\bibinfo{year}{2016}), \eprint{doi:10.1038/natrevmats.2016.50}.

\bibitem[{\citenamefont{Sothmann et~al.}(2015)\citenamefont{Sothmann, S{\'a}nchez, and Jordan}}]{Sothman2015}
\bibinfo{author}{\bibfnamefont{B.}~\bibnamefont{Sothmann}}, \bibinfo{author}{\bibfnamefont{R.}~\bibnamefont{S{\'a}nchez}}, \bibnamefont{and} \bibinfo{author}{\bibfnamefont{A.~N.} \bibnamefont{Jordan}}, \bibinfo{journal}{Nanotechnology} \textbf{\bibinfo{volume}{26}}, \bibinfo{pages}{032001} (\bibinfo{year}{2015}), \eprint{http://stacks.iop.org/0957-4484/26/i=3/a=032001}.

\bibitem[{\citenamefont{Reddy et~al.}(2007)\citenamefont{Reddy, Jang, Segalman, and Majumdar}}]{Reddy2007}
\bibinfo{author}{\bibfnamefont{P.}~\bibnamefont{Reddy}}, \bibinfo{author}{\bibfnamefont{S.-Y.} \bibnamefont{Jang}}, \bibinfo{author}{\bibfnamefont{R.~A.} \bibnamefont{Segalman}}, \bibnamefont{and} \bibinfo{author}{\bibfnamefont{A.}~\bibnamefont{Majumdar}}, \bibinfo{journal}{Science} \textbf{\bibinfo{volume}{315}}, \bibinfo{pages}{1568} (\bibinfo{year}{2007}), \eprint{doi:10.1126/science.1137149}.

\bibitem[{\citenamefont{Li et~al.}(2021)\citenamefont{Li, Li, Han, Zheng, Li, Li, Fan, and Qiu}}]{Li2021transformingmeta}
\bibinfo{author}{\bibfnamefont{Y.}~\bibnamefont{Li}}, \bibinfo{author}{\bibfnamefont{W.}~\bibnamefont{Li}}, \bibinfo{author}{\bibfnamefont{T.}~\bibnamefont{Han}}, \bibinfo{author}{\bibfnamefont{X.}~\bibnamefont{Zheng}}, \bibinfo{author}{\bibfnamefont{J.}~\bibnamefont{Li}}, \bibinfo{author}{\bibfnamefont{B.}~\bibnamefont{Li}}, \bibinfo{author}{\bibfnamefont{S.}~\bibnamefont{Fan}}, \bibnamefont{and} \bibinfo{author}{\bibfnamefont{C.-W.} \bibnamefont{Qiu}}, \bibinfo{journal}{Nature Reviews Materials} \textbf{\bibinfo{volume}{6}}, \bibinfo{pages}{488} (\bibinfo{year}{2021}), \eprint{doi:10.1038/s41578-021-00283-2}.

\bibitem[{\citenamefont{Yang et~al.}(2024)\citenamefont{Yang, Zhang, Xu, Liu, Jin, Zhuang, Lei, Liu, Jiang, Ouyang et~al.}}]{Yang2024metamaterials}
\bibinfo{author}{\bibfnamefont{F.}~\bibnamefont{Yang}}, \bibinfo{author}{\bibfnamefont{Z.}~\bibnamefont{Zhang}}, \bibinfo{author}{\bibfnamefont{L.}~\bibnamefont{Xu}}, \bibinfo{author}{\bibfnamefont{Z.}~\bibnamefont{Liu}}, \bibinfo{author}{\bibfnamefont{P.}~\bibnamefont{Jin}}, \bibinfo{author}{\bibfnamefont{P.}~\bibnamefont{Zhuang}}, \bibinfo{author}{\bibfnamefont{M.}~\bibnamefont{Lei}}, \bibinfo{author}{\bibfnamefont{J.}~\bibnamefont{Liu}}, \bibinfo{author}{\bibfnamefont{J.-H.} \bibnamefont{Jiang}}, \bibinfo{author}{\bibfnamefont{X.}~\bibnamefont{Ouyang}}, \bibnamefont{et~al.}, \bibinfo{journal}{Rev. Mod. Phys.} \textbf{\bibinfo{volume}{96}}, \bibinfo{pages}{015002} (\bibinfo{year}{2024}).

\bibitem[{\citenamefont{Reimann}(2002)}]{Reimann2002}
\bibinfo{author}{\bibfnamefont{P.}~\bibnamefont{Reimann}}, \bibinfo{journal}{Phys. Rev.} \textbf{\bibinfo{volume}{361}}, \bibinfo{pages}{57} (\bibinfo{year}{2002}), ISSN \bibinfo{issn}{0370-1573}, \eprint{doi:10.1016/S0370-1573(01)00081-3}.

\bibitem[{\citenamefont{Brey and Casado}(1990)}]{brey1990generalized}
\bibinfo{author}{\bibfnamefont{J.}~\bibnamefont{Brey}} \bibnamefont{and} \bibinfo{author}{\bibfnamefont{J.}~\bibnamefont{Casado}}, \bibinfo{journal}{J. Stat. Phys.} \textbf{\bibinfo{volume}{61}}, \bibinfo{pages}{713} (\bibinfo{year}{1990}).

\bibitem[{\citenamefont{Popov and Hernandez}(2007)}]{hern07a}
\bibinfo{author}{\bibfnamefont{A.~V.} \bibnamefont{Popov}} \bibnamefont{and} \bibinfo{author}{\bibfnamefont{R.}~\bibnamefont{Hernandez}}, \bibinfo{journal}{J. Chem. Phys.} \textbf{\bibinfo{volume}{126}}, \bibinfo{pages}{244506} (\bibinfo{year}{2007}), \eprint{doi:10.1063/1.2743032}.

\bibitem[{\citenamefont{Popov and Hernandez}(2013)}]{hern13d}
\bibinfo{author}{\bibfnamefont{A.~V.} \bibnamefont{Popov}} \bibnamefont{and} \bibinfo{author}{\bibfnamefont{R.}~\bibnamefont{Hernandez}}, \bibinfo{journal}{Phys. Rev. E} \textbf{\bibinfo{volume}{88}}, \bibinfo{pages}{032145} (\bibinfo{year}{2013}), \eprint{10.1103/PhysRevE.88.032145}.

\bibitem[{\citenamefont{Ford et~al.}(2015)\citenamefont{Ford, Laker, and Charlesworth}}]{Ford2015}
\bibinfo{author}{\bibfnamefont{I.~J.} \bibnamefont{Ford}}, \bibinfo{author}{\bibfnamefont{Z.~P.~L.} \bibnamefont{Laker}}, \bibnamefont{and} \bibinfo{author}{\bibfnamefont{H.~J.} \bibnamefont{Charlesworth}}, \bibinfo{journal}{Phys. Rev. E} \textbf{\bibinfo{volume}{92}}, \bibinfo{pages}{042108} (\bibinfo{year}{2015}), \eprint{doi:10.1103/PhysRevE.92.042108}.

\bibitem[{\citenamefont{Brandner et~al.}(2015)\citenamefont{Brandner, Saito, and Seifert}}]{Seifert2015periodictemp}
\bibinfo{author}{\bibfnamefont{K.}~\bibnamefont{Brandner}}, \bibinfo{author}{\bibfnamefont{K.}~\bibnamefont{Saito}}, \bibnamefont{and} \bibinfo{author}{\bibfnamefont{U.}~\bibnamefont{Seifert}}, \bibinfo{journal}{Phys. Rev. X} \textbf{\bibinfo{volume}{5}}, \bibinfo{pages}{031019} (\bibinfo{year}{2015}), \eprint{doi:10.1103/PhysRevX.5.031019}.

\bibitem[{\citenamefont{Brandner and Seifert}(2016)}]{Seifert2016periodiccurrent}
\bibinfo{author}{\bibfnamefont{K.}~\bibnamefont{Brandner}} \bibnamefont{and} \bibinfo{author}{\bibfnamefont{U.}~\bibnamefont{Seifert}}, \bibinfo{journal}{Phys. Rev. E} \textbf{\bibinfo{volume}{93}}, \bibinfo{pages}{062134} (\bibinfo{year}{2016}), \eprint{doi:10.1103/PhysRevE.93.062134}.

\bibitem[{\citenamefont{Awasthi and Dutta}(2021)}]{Awasthi2021}
\bibinfo{author}{\bibfnamefont{S.}~\bibnamefont{Awasthi}} \bibnamefont{and} \bibinfo{author}{\bibfnamefont{S.~B.} \bibnamefont{Dutta}}, \bibinfo{journal}{Phys. Rev. E} \textbf{\bibinfo{volume}{103}}, \bibinfo{pages}{062143} (\bibinfo{year}{2021}), \eprint{doi:10.1103/PhysRevE.103.062143}.

\bibitem[{\citenamefont{Portugal et~al.}(2022)\citenamefont{Portugal, Brange, and Flindt}}]{Portugal2022effective}
\bibinfo{author}{\bibfnamefont{P.}~\bibnamefont{Portugal}}, \bibinfo{author}{\bibfnamefont{F.}~\bibnamefont{Brange}}, \bibnamefont{and} \bibinfo{author}{\bibfnamefont{C.}~\bibnamefont{Flindt}}, \bibinfo{journal}{Phys. Rev. Res.} \textbf{\bibinfo{volume}{4}}, \bibinfo{pages}{043112} (\bibinfo{year}{2022}), \eprint{doi:10.1103/PhysRevResearch.4.043112}.

\bibitem[{\citenamefont{Ordonez-Miranda et~al.}(2022)\citenamefont{Ordonez-Miranda, Anufriev, Nomura, and Volz}}]{Volz2022}
\bibinfo{author}{\bibfnamefont{J.}~\bibnamefont{Ordonez-Miranda}}, \bibinfo{author}{\bibfnamefont{R.}~\bibnamefont{Anufriev}}, \bibinfo{author}{\bibfnamefont{M.}~\bibnamefont{Nomura}}, \bibnamefont{and} \bibinfo{author}{\bibfnamefont{S.}~\bibnamefont{Volz}}, \bibinfo{journal}{Phys. Rev. B} \textbf{\bibinfo{volume}{106}}, \bibinfo{pages}{L100102} (\bibinfo{year}{2022}), \eprint{doi:10.1103/PhysRevB.106.L100102}.

\bibitem[{\citenamefont{Lanoisel\'ee et~al.}(2022)\citenamefont{Lanoisel\'ee, Stanislavsky, Calebiro, and Weron}}]{Weron2022}
\bibinfo{author}{\bibfnamefont{Y.}~\bibnamefont{Lanoisel\'ee}}, \bibinfo{author}{\bibfnamefont{A.}~\bibnamefont{Stanislavsky}}, \bibinfo{author}{\bibfnamefont{D.}~\bibnamefont{Calebiro}}, \bibnamefont{and} \bibinfo{author}{\bibfnamefont{A.}~\bibnamefont{Weron}}, \bibinfo{journal}{Phys. Rev. E} \textbf{\bibinfo{volume}{106}}, \bibinfo{pages}{064127} (\bibinfo{year}{2022}), \eprint{doi:10.1103/PhysRevE.106.064127}.

\bibitem[{\citenamefont{Ben-Abdallah}(2017)}]{Ben-Abdallah2017thermalmemristor}
\bibinfo{author}{\bibfnamefont{P.}~\bibnamefont{Ben-Abdallah}}, \bibinfo{journal}{AIP Adv.} \textbf{\bibinfo{volume}{7}}, \bibinfo{pages}{065002} (\bibinfo{year}{2017}), \eprint{doi:10.1063/1.4985055}.

\bibitem[{\citenamefont{Ordonez-Miranda et~al.}(2019)\citenamefont{Ordonez-Miranda, Ezzahri, Tiburcio-Moreno, Joulain, and Drevillon}}]{Ordonez-Miranda2019thermalmemristor}
\bibinfo{author}{\bibfnamefont{J.}~\bibnamefont{Ordonez-Miranda}}, \bibinfo{author}{\bibfnamefont{Y.}~\bibnamefont{Ezzahri}}, \bibinfo{author}{\bibfnamefont{J.~A.} \bibnamefont{Tiburcio-Moreno}}, \bibinfo{author}{\bibfnamefont{K.}~\bibnamefont{Joulain}}, \bibnamefont{and} \bibinfo{author}{\bibfnamefont{J.}~\bibnamefont{Drevillon}}, \bibinfo{journal}{Phys. Rev. Lett.} \textbf{\bibinfo{volume}{123}}, \bibinfo{pages}{025901} (\bibinfo{year}{2019}), \eprint{doi:10.1103/PhysRevLett.123.025901}.

\bibitem[{\citenamefont{Kuzkin and Krivtsov}(2020)}]{Krivtsov2020}
\bibinfo{author}{\bibfnamefont{V.~A.} \bibnamefont{Kuzkin}} \bibnamefont{and} \bibinfo{author}{\bibfnamefont{A.~M.} \bibnamefont{Krivtsov}}, \bibinfo{journal}{Phys. Rev. E} \textbf{\bibinfo{volume}{101}}, \bibinfo{pages}{042209} (\bibinfo{year}{2020}), \eprint{doi:10.1103/PhysRevE.101.042209}.

\bibitem[{\citenamefont{Ochoa}(2024)}]{Ochoa2024}
\bibinfo{author}{\bibfnamefont{M.~A.} \bibnamefont{Ochoa}}, \bibinfo{journal}{Phys. Rev. B} \textbf{\bibinfo{volume}{110}}, \bibinfo{pages}{155411} (\bibinfo{year}{2024}), \eprint{doi:10.1103/PhysRevB.110.155411}.

\bibitem[{\citenamefont{Ochoa}(2025)}]{ochoa2025semiclassical}
\bibinfo{author}{\bibfnamefont{M.~A.} \bibnamefont{Ochoa}}, \bibinfo{journal}{arXiv preprint arXiv:2503.23620}  (\bibinfo{year}{2025}).

\bibitem[{\citenamefont{Migliore and Nitzan}(2013)}]{Migliore2013}
\bibinfo{author}{\bibfnamefont{A.}~\bibnamefont{Migliore}} \bibnamefont{and} \bibinfo{author}{\bibfnamefont{A.}~\bibnamefont{Nitzan}}, \bibinfo{journal}{J. Am. Chem. Soc.} \textbf{\bibinfo{volume}{135}}, \bibinfo{pages}{9420} (\bibinfo{year}{2013}), \eprint{doi:10.1021/ja401336u}.

\bibitem[{\citenamefont{Galperin et~al.}(2005)\citenamefont{Galperin, Ratner, and Nitzan}}]{galperin2005hysteresis}
\bibinfo{author}{\bibfnamefont{M.}~\bibnamefont{Galperin}}, \bibinfo{author}{\bibfnamefont{M.~A.} \bibnamefont{Ratner}}, \bibnamefont{and} \bibinfo{author}{\bibfnamefont{A.}~\bibnamefont{Nitzan}}, \bibinfo{journal}{Nano Lett.} \textbf{\bibinfo{volume}{5}}, \bibinfo{pages}{125} (\bibinfo{year}{2005}).

\bibitem[{\citenamefont{Chen et~al.}(2023)\citenamefont{Chen, Gibson, and Craven}}]{craven2023b}
\bibinfo{author}{\bibfnamefont{R.}~\bibnamefont{Chen}}, \bibinfo{author}{\bibfnamefont{T.}~\bibnamefont{Gibson}}, \bibnamefont{and} \bibinfo{author}{\bibfnamefont{G.~T.} \bibnamefont{Craven}}, \bibinfo{journal}{Phys. Rev. E} \textbf{\bibinfo{volume}{108}}, \bibinfo{pages}{024148} (\bibinfo{year}{2023}), \eprint{doi:10.1103/PhysRevE.108.024148}.

\bibitem[{\citenamefont{Chen et~al.}(2024)\citenamefont{Chen, Gibson, and Craven}}]{craven2024a}
\bibinfo{author}{\bibfnamefont{R.}~\bibnamefont{Chen}}, \bibinfo{author}{\bibfnamefont{T.}~\bibnamefont{Gibson}}, \bibnamefont{and} \bibinfo{author}{\bibfnamefont{G.~T.} \bibnamefont{Craven}}, \bibinfo{journal}{J. Chem. Phys.} \textbf{\bibinfo{volume}{160}}, \bibinfo{pages}{194305} (\bibinfo{year}{2024}), \eprint{doi:10.1063/5.0204819}.

\bibitem[{\citenamefont{Chen and Craven}(2024)}]{Chen2024}
\bibinfo{author}{\bibfnamefont{R.}~\bibnamefont{Chen}} \bibnamefont{and} \bibinfo{author}{\bibfnamefont{G.~T.} \bibnamefont{Craven}}, \bibinfo{journal}{Journal of Physics: Condensed Matter} \textbf{\bibinfo{volume}{36}}, \bibinfo{pages}{405201} (\bibinfo{year}{2024}), \eprint{doi:10.1088/1361-648X/ad5d40}.

\bibitem[{\citenamefont{Chen and Craven}(2025)}]{craven2025entropy}
\bibinfo{author}{\bibfnamefont{R.}~\bibnamefont{Chen}} \bibnamefont{and} \bibinfo{author}{\bibfnamefont{G.~T.} \bibnamefont{Craven}}, \bibinfo{journal}{Entropy} \textbf{\bibinfo{volume}{27}} (\bibinfo{year}{2025}).

\bibitem[{\citenamefont{Galperin et~al.}(2007)\citenamefont{Galperin, Nitzan, and Ratner}}]{Nitzan2007}
\bibinfo{author}{\bibfnamefont{M.}~\bibnamefont{Galperin}}, \bibinfo{author}{\bibfnamefont{A.}~\bibnamefont{Nitzan}}, \bibnamefont{and} \bibinfo{author}{\bibfnamefont{M.~A.} \bibnamefont{Ratner}}, \bibinfo{journal}{Phys. Rev. B} \textbf{\bibinfo{volume}{75}}, \bibinfo{pages}{155312} (\bibinfo{year}{2007}), \eprint{doi:10.1103/PhysRevB.75.155312}.

\bibitem[{\citenamefont{Leitner}(2008)}]{Leitner2008}
\bibinfo{author}{\bibfnamefont{D.~M.} \bibnamefont{Leitner}}, \bibinfo{journal}{Annu. Rev. Phys. Chem.} \textbf{\bibinfo{volume}{59}}, \bibinfo{pages}{233} (\bibinfo{year}{2008}), \eprint{doi:10.1146/annurev.physchem.59.032607.093606}.

\bibitem[{\citenamefont{Leitner}(2013)}]{Leitner2013}
\bibinfo{author}{\bibfnamefont{D.~M.} \bibnamefont{Leitner}}, \bibinfo{journal}{J. Phys. Chem. B} \textbf{\bibinfo{volume}{117}}, \bibinfo{pages}{12820} (\bibinfo{year}{2013}), \eprint{doi:10.1021/jp402012z}.

\bibitem[{\citenamefont{Lim et~al.}(2013)\citenamefont{Lim, L\'opez, and S\'anchez}}]{Lim2013}
\bibinfo{author}{\bibfnamefont{J.~S.} \bibnamefont{Lim}}, \bibinfo{author}{\bibfnamefont{R.}~\bibnamefont{L\'opez}}, \bibnamefont{and} \bibinfo{author}{\bibfnamefont{D.}~\bibnamefont{S\'anchez}}, \bibinfo{journal}{Phys. Rev. B} \textbf{\bibinfo{volume}{88}}, \bibinfo{pages}{201304} (\bibinfo{year}{2013}), \eprint{doi:10.1103/PhysRevB.88.201304}.

\bibitem[{\citenamefont{Zhu et~al.}(2021)\citenamefont{Zhu, Liu, and He}}]{He2021}
\bibinfo{author}{\bibfnamefont{J.}~\bibnamefont{Zhu}}, \bibinfo{author}{\bibfnamefont{Y.}~\bibnamefont{Liu}}, \bibnamefont{and} \bibinfo{author}{\bibfnamefont{D.}~\bibnamefont{He}}, \bibinfo{journal}{Phys. Rev. E} \textbf{\bibinfo{volume}{103}}, \bibinfo{pages}{062121} (\bibinfo{year}{2021}), \eprint{doi:10.1103/PhysRevE.103.062121}.

\bibitem[{\citenamefont{Wei and Hernandez}(0)}]{HernandezJPCL2023}
\bibinfo{author}{\bibfnamefont{X.}~\bibnamefont{Wei}} \bibnamefont{and} \bibinfo{author}{\bibfnamefont{R.}~\bibnamefont{Hernandez}}, \bibinfo{journal}{J. Phys. Chem. Lett.} \textbf{\bibinfo{volume}{0}}, \bibinfo{pages}{9834} (\bibinfo{year}{0}), \eprint{doi:10.1021/acs.jpclett.3c02367}.

\bibitem[{\citenamefont{Sharony et~al.}(2020)\citenamefont{Sharony, Chen, and Nitzan}}]{sharony2020stochastic}
\bibinfo{author}{\bibfnamefont{I.}~\bibnamefont{Sharony}}, \bibinfo{author}{\bibfnamefont{R.}~\bibnamefont{Chen}}, \bibnamefont{and} \bibinfo{author}{\bibfnamefont{A.}~\bibnamefont{Nitzan}}, \bibinfo{journal}{J. Chem. Phys.} \textbf{\bibinfo{volume}{153}}, \bibinfo{pages}{144113} (\bibinfo{year}{2020}), \eprint{doi:10.1063/5.0022423}.

\bibitem[{\citenamefont{Dmitriev et~al.}(2023)\citenamefont{Dmitriev, Kuzkin, and Krivtsov}}]{Krivtsov2023}
\bibinfo{author}{\bibfnamefont{S.~V.} \bibnamefont{Dmitriev}}, \bibinfo{author}{\bibfnamefont{V.~A.} \bibnamefont{Kuzkin}}, \bibnamefont{and} \bibinfo{author}{\bibfnamefont{A.~M.} \bibnamefont{Krivtsov}}, \bibinfo{journal}{Phys. Rev. E} \textbf{\bibinfo{volume}{108}}, \bibinfo{pages}{054221} (\bibinfo{year}{2023}), \eprint{doi:10.1103/PhysRevE.108.054221}.

\bibitem[{\citenamefont{Tan et~al.}(2011)\citenamefont{Tan, Balachandran, Sadat, Gavini, Dunietz, Jang, and Reddy}}]{Tan2011}
\bibinfo{author}{\bibfnamefont{A.}~\bibnamefont{Tan}}, \bibinfo{author}{\bibfnamefont{J.}~\bibnamefont{Balachandran}}, \bibinfo{author}{\bibfnamefont{S.}~\bibnamefont{Sadat}}, \bibinfo{author}{\bibfnamefont{V.}~\bibnamefont{Gavini}}, \bibinfo{author}{\bibfnamefont{B.~D.} \bibnamefont{Dunietz}}, \bibinfo{author}{\bibfnamefont{S.-Y.} \bibnamefont{Jang}}, \bibnamefont{and} \bibinfo{author}{\bibfnamefont{P.}~\bibnamefont{Reddy}}, \bibinfo{journal}{J. Am. Chem. Soc.} \textbf{\bibinfo{volume}{133}}, \bibinfo{pages}{8838} (\bibinfo{year}{2011}), \eprint{doi:10.1021/ja202178k}.

\bibitem[{\citenamefont{Lee et~al.}(2013)\citenamefont{Lee, Kim, Jeong, Zotti, Pauly, Cuevas, and Reddy}}]{Lee2013}
\bibinfo{author}{\bibfnamefont{W.}~\bibnamefont{Lee}}, \bibinfo{author}{\bibfnamefont{K.}~\bibnamefont{Kim}}, \bibinfo{author}{\bibfnamefont{W.}~\bibnamefont{Jeong}}, \bibinfo{author}{\bibfnamefont{L.~A.} \bibnamefont{Zotti}}, \bibinfo{author}{\bibfnamefont{F.}~\bibnamefont{Pauly}}, \bibinfo{author}{\bibfnamefont{J.~C.} \bibnamefont{Cuevas}}, \bibnamefont{and} \bibinfo{author}{\bibfnamefont{P.}~\bibnamefont{Reddy}}, \bibinfo{journal}{Nature} \textbf{\bibinfo{volume}{498}}, \bibinfo{pages}{209} (\bibinfo{year}{2013}), \eprint{doi:10.1038/nature12183}.

\bibitem[{\citenamefont{Kim et~al.}(2014)\citenamefont{Kim, Jeong, Kim, Lee, and Reddy}}]{Kim2014}
\bibinfo{author}{\bibfnamefont{Y.}~\bibnamefont{Kim}}, \bibinfo{author}{\bibfnamefont{W.}~\bibnamefont{Jeong}}, \bibinfo{author}{\bibfnamefont{K.}~\bibnamefont{Kim}}, \bibinfo{author}{\bibfnamefont{W.}~\bibnamefont{Lee}}, \bibnamefont{and} \bibinfo{author}{\bibfnamefont{P.}~\bibnamefont{Reddy}}, \bibinfo{journal}{Nature Nanotech.} \textbf{\bibinfo{volume}{9}}, \bibinfo{pages}{881} (\bibinfo{year}{2014}), \eprint{doi:10.1038/nnano.2014.209}.

\bibitem[{\citenamefont{Capozzi et~al.}(2015)\citenamefont{Capozzi, Xia, Adak, Dell, Liu, Taylor, Neaton, Campos, and Venkataraman}}]{Venkataraman2015}
\bibinfo{author}{\bibfnamefont{B.}~\bibnamefont{Capozzi}}, \bibinfo{author}{\bibfnamefont{J.}~\bibnamefont{Xia}}, \bibinfo{author}{\bibfnamefont{O.}~\bibnamefont{Adak}}, \bibinfo{author}{\bibfnamefont{E.~J.} \bibnamefont{Dell}}, \bibinfo{author}{\bibfnamefont{Z.-F.} \bibnamefont{Liu}}, \bibinfo{author}{\bibfnamefont{J.~C.} \bibnamefont{Taylor}}, \bibinfo{author}{\bibfnamefont{J.~B.} \bibnamefont{Neaton}}, \bibinfo{author}{\bibfnamefont{L.~M.} \bibnamefont{Campos}}, \bibnamefont{and} \bibinfo{author}{\bibfnamefont{L.}~\bibnamefont{Venkataraman}}, \bibinfo{journal}{Nature Nanotech.} \textbf{\bibinfo{volume}{10}}, \bibinfo{pages}{522} (\bibinfo{year}{2015}), \eprint{doi:10.1038/nnano.2015.97}.

\bibitem[{\citenamefont{Garner et~al.}(2018)\citenamefont{Garner, Li, Chen, Su, Shangguan, Paley, Liu, Ng, Li, Xiao et~al.}}]{Garner2018}
\bibinfo{author}{\bibfnamefont{M.~H.} \bibnamefont{Garner}}, \bibinfo{author}{\bibfnamefont{H.}~\bibnamefont{Li}}, \bibinfo{author}{\bibfnamefont{Y.}~\bibnamefont{Chen}}, \bibinfo{author}{\bibfnamefont{T.~A.} \bibnamefont{Su}}, \bibinfo{author}{\bibfnamefont{Z.}~\bibnamefont{Shangguan}}, \bibinfo{author}{\bibfnamefont{D.~W.} \bibnamefont{Paley}}, \bibinfo{author}{\bibfnamefont{T.}~\bibnamefont{Liu}}, \bibinfo{author}{\bibfnamefont{F.}~\bibnamefont{Ng}}, \bibinfo{author}{\bibfnamefont{H.}~\bibnamefont{Li}}, \bibinfo{author}{\bibfnamefont{S.}~\bibnamefont{Xiao}}, \bibnamefont{et~al.}, \bibinfo{journal}{Nature} \textbf{\bibinfo{volume}{558}}, \bibinfo{pages}{415} (\bibinfo{year}{2018}), \eprint{doi:10.1038/s41586-018-0197-9}.

\bibitem[{\citenamefont{Cui et~al.}(2019)\citenamefont{Cui, Hur, Akbar, Kl\"ockner, Jeong, Pauly, Jang, Reddy, and Meyhofer}}]{Reddy2019nature}
\bibinfo{author}{\bibfnamefont{L.}~\bibnamefont{Cui}}, \bibinfo{author}{\bibfnamefont{S.}~\bibnamefont{Hur}}, \bibinfo{author}{\bibfnamefont{Z.~A.} \bibnamefont{Akbar}}, \bibinfo{author}{\bibfnamefont{J.~C.} \bibnamefont{Kl\"ockner}}, \bibinfo{author}{\bibfnamefont{W.}~\bibnamefont{Jeong}}, \bibinfo{author}{\bibfnamefont{F.}~\bibnamefont{Pauly}}, \bibinfo{author}{\bibfnamefont{S.-Y.} \bibnamefont{Jang}}, \bibinfo{author}{\bibfnamefont{P.}~\bibnamefont{Reddy}}, \bibnamefont{and} \bibinfo{author}{\bibfnamefont{E.}~\bibnamefont{Meyhofer}}, \bibinfo{journal}{Nature} \textbf{\bibinfo{volume}{572}}, \bibinfo{pages}{628} (\bibinfo{year}{2019}), \eprint{doi:10.1038/s41586-019-1420-z}.

\bibitem[{\citenamefont{Mosso et~al.}(2019)\citenamefont{Mosso, Sadeghi, Gemma, Sangtarash, Drechsler, Lambert, and Gotsmann}}]{Mosso2019}
\bibinfo{author}{\bibfnamefont{N.}~\bibnamefont{Mosso}}, \bibinfo{author}{\bibfnamefont{H.}~\bibnamefont{Sadeghi}}, \bibinfo{author}{\bibfnamefont{A.}~\bibnamefont{Gemma}}, \bibinfo{author}{\bibfnamefont{S.}~\bibnamefont{Sangtarash}}, \bibinfo{author}{\bibfnamefont{U.}~\bibnamefont{Drechsler}}, \bibinfo{author}{\bibfnamefont{C.}~\bibnamefont{Lambert}}, \bibnamefont{and} \bibinfo{author}{\bibfnamefont{B.}~\bibnamefont{Gotsmann}}, \bibinfo{journal}{Nano Letters} \textbf{\bibinfo{volume}{19}}, \bibinfo{pages}{7614} (\bibinfo{year}{2019}), \eprint{doi:10.1021/acs.nanolett.9b02089}.

\bibitem[{\citenamefont{Zimbovskaya and Nitzan}(2020)}]{Zimbovskaya2020}
\bibinfo{author}{\bibfnamefont{N.~A.} \bibnamefont{Zimbovskaya}} \bibnamefont{and} \bibinfo{author}{\bibfnamefont{A.}~\bibnamefont{Nitzan}}, \bibinfo{journal}{J. Phys. Chem. B} \textbf{\bibinfo{volume}{124}}, \bibinfo{pages}{2632} (\bibinfo{year}{2020}), \eprint{doi:10.1021/acs.jpcb.0c00059}.

\bibitem[{\citenamefont{Nitzan and Ratner}(2003)}]{Nitzan2003electron}
\bibinfo{author}{\bibfnamefont{A.}~\bibnamefont{Nitzan}} \bibnamefont{and} \bibinfo{author}{\bibfnamefont{M.~A.} \bibnamefont{Ratner}}, \bibinfo{journal}{Science} \textbf{\bibinfo{volume}{300}}, \bibinfo{pages}{1384} (\bibinfo{year}{2003}), \eprint{doi:10.1126/science.1081572}.

\bibitem[{\citenamefont{Li et~al.}(2006)\citenamefont{Li, Wang, and Casati}}]{Li2006}
\bibinfo{author}{\bibfnamefont{B.}~\bibnamefont{Li}}, \bibinfo{author}{\bibfnamefont{L.}~\bibnamefont{Wang}}, \bibnamefont{and} \bibinfo{author}{\bibfnamefont{G.}~\bibnamefont{Casati}}, \bibinfo{journal}{Appl. Phys. Lett.} \textbf{\bibinfo{volume}{88}}, \bibinfo{pages}{143501} (\bibinfo{year}{2006}), \eprint{doi:10.1063/1.2191730}.

\bibitem[{\citenamefont{Ben-Abdallah and Biehs}(2014)}]{Ben-Abdallah2014}
\bibinfo{author}{\bibfnamefont{P.}~\bibnamefont{Ben-Abdallah}} \bibnamefont{and} \bibinfo{author}{\bibfnamefont{S.-A.} \bibnamefont{Biehs}}, \bibinfo{journal}{Phys. Rev. Lett.} \textbf{\bibinfo{volume}{112}}, \bibinfo{pages}{044301} (\bibinfo{year}{2014}), \eprint{doi:10.1103/PhysRevLett.112.044301}.

\bibitem[{\citenamefont{Wang et~al.}(2017)\citenamefont{Wang, Cottrill, Kunai, Toland, Liu, Wang, and Strano}}]{Wang2017thermaldiode}
\bibinfo{author}{\bibfnamefont{S.}~\bibnamefont{Wang}}, \bibinfo{author}{\bibfnamefont{A.~L.} \bibnamefont{Cottrill}}, \bibinfo{author}{\bibfnamefont{Y.}~\bibnamefont{Kunai}}, \bibinfo{author}{\bibfnamefont{A.~R.} \bibnamefont{Toland}}, \bibinfo{author}{\bibfnamefont{P.}~\bibnamefont{Liu}}, \bibinfo{author}{\bibfnamefont{W.-J.} \bibnamefont{Wang}}, \bibnamefont{and} \bibinfo{author}{\bibfnamefont{M.~S.} \bibnamefont{Strano}}, \bibinfo{journal}{Phys. Chem. Chem. Phys.} \textbf{\bibinfo{volume}{19}}, \bibinfo{pages}{13172} (\bibinfo{year}{2017}), \eprint{doi:10.1039/C7CP02445B}.

\bibitem[{\citenamefont{Wang and Li}(2007)}]{Li2007logic}
\bibinfo{author}{\bibfnamefont{L.}~\bibnamefont{Wang}} \bibnamefont{and} \bibinfo{author}{\bibfnamefont{B.}~\bibnamefont{Li}}, \bibinfo{journal}{Phys. Rev. Lett.} \textbf{\bibinfo{volume}{99}}, \bibinfo{pages}{177208} (\bibinfo{year}{2007}), \eprint{doi:10.1103/PhysRevLett.99.177208}.

\bibitem[{\citenamefont{van De~Burgt et~al.}(2018)\citenamefont{van De~Burgt, Melianas, Keene, Malliaras, and Salleo}}]{Van2018organic}
\bibinfo{author}{\bibfnamefont{Y.}~\bibnamefont{van De~Burgt}}, \bibinfo{author}{\bibfnamefont{A.}~\bibnamefont{Melianas}}, \bibinfo{author}{\bibfnamefont{S.~T.} \bibnamefont{Keene}}, \bibinfo{author}{\bibfnamefont{G.}~\bibnamefont{Malliaras}}, \bibnamefont{and} \bibinfo{author}{\bibfnamefont{A.}~\bibnamefont{Salleo}}, \bibinfo{journal}{Nature electronics} \textbf{\bibinfo{volume}{1}}, \bibinfo{pages}{386} (\bibinfo{year}{2018}).

\bibitem[{\citenamefont{Markovi{\'c} et~al.}(2020)\citenamefont{Markovi{\'c}, Mizrahi, Querlioz, and Grollier}}]{markovic2020physics}
\bibinfo{author}{\bibfnamefont{D.}~\bibnamefont{Markovi{\'c}}}, \bibinfo{author}{\bibfnamefont{A.}~\bibnamefont{Mizrahi}}, \bibinfo{author}{\bibfnamefont{D.}~\bibnamefont{Querlioz}}, \bibnamefont{and} \bibinfo{author}{\bibfnamefont{J.}~\bibnamefont{Grollier}}, \bibinfo{journal}{Nature Reviews Physics} \textbf{\bibinfo{volume}{2}}, \bibinfo{pages}{499} (\bibinfo{year}{2020}).

\bibitem[{\citenamefont{Kudithipudi et~al.}(2025)\citenamefont{Kudithipudi, Schuman, Vineyard, Pandit, Merkel, Kubendran, Aimone, Orchard, Mayr, Benosman et~al.}}]{kudithipudi2025neuromorphic}
\bibinfo{author}{\bibfnamefont{D.}~\bibnamefont{Kudithipudi}}, \bibinfo{author}{\bibfnamefont{C.}~\bibnamefont{Schuman}}, \bibinfo{author}{\bibfnamefont{C.~M.} \bibnamefont{Vineyard}}, \bibinfo{author}{\bibfnamefont{T.}~\bibnamefont{Pandit}}, \bibinfo{author}{\bibfnamefont{C.}~\bibnamefont{Merkel}}, \bibinfo{author}{\bibfnamefont{R.}~\bibnamefont{Kubendran}}, \bibinfo{author}{\bibfnamefont{J.~B.} \bibnamefont{Aimone}}, \bibinfo{author}{\bibfnamefont{G.}~\bibnamefont{Orchard}}, \bibinfo{author}{\bibfnamefont{C.}~\bibnamefont{Mayr}}, \bibinfo{author}{\bibfnamefont{R.}~\bibnamefont{Benosman}}, \bibnamefont{et~al.}, \bibinfo{journal}{Nature} \textbf{\bibinfo{volume}{637}}, \bibinfo{pages}{801} (\bibinfo{year}{2025}).

\bibitem[{\citenamefont{Caravelli and Carbajal}(2018)}]{Caravelli2018}
\bibinfo{author}{\bibfnamefont{F.}~\bibnamefont{Caravelli}} \bibnamefont{and} \bibinfo{author}{\bibfnamefont{J.~P.} \bibnamefont{Carbajal}}, \bibinfo{journal}{Technologies} \textbf{\bibinfo{volume}{6}} (\bibinfo{year}{2018}), \eprint{doi:10.3390/technologies6040118}.

\bibitem[{\citenamefont{Sangwan and Hersam}(2020)}]{Sangwan2020neuromorphic}
\bibinfo{author}{\bibfnamefont{V.~K.} \bibnamefont{Sangwan}} \bibnamefont{and} \bibinfo{author}{\bibfnamefont{M.~C.} \bibnamefont{Hersam}}, \bibinfo{journal}{Nature Nanotech.} \textbf{\bibinfo{volume}{15}}, \bibinfo{pages}{517} (\bibinfo{year}{2020}), \eprint{doi:10.1038/s41565-020-0647-z}.

\bibitem[{\citenamefont{Pershin and Ventra}(2011)}]{Yuriy2011}
\bibinfo{author}{\bibfnamefont{Y.~V.} \bibnamefont{Pershin}} \bibnamefont{and} \bibinfo{author}{\bibfnamefont{M.~D.} \bibnamefont{Ventra}}, \bibinfo{journal}{Adv. Phys.} \textbf{\bibinfo{volume}{60}}, \bibinfo{pages}{145} (\bibinfo{year}{2011}), \eprint{doi:10.1080/00018732.2010.544961}.

\bibitem[{\citenamefont{Jorgensen et~al.}(1996)\citenamefont{Jorgensen, Maxwell, and Tirado-Rives}}]{jorgensen1996development}
\bibinfo{author}{\bibfnamefont{W.~L.} \bibnamefont{Jorgensen}}, \bibinfo{author}{\bibfnamefont{D.~S.} \bibnamefont{Maxwell}}, \bibnamefont{and} \bibinfo{author}{\bibfnamefont{J.}~\bibnamefont{Tirado-Rives}}, \bibinfo{journal}{J. Am. Chem. Soc.} \textbf{\bibinfo{volume}{118}}, \bibinfo{pages}{11225} (\bibinfo{year}{1996}).

\bibitem[{\citenamefont{Sekimoto}(1998)}]{Sekimoto1998}
\bibinfo{author}{\bibfnamefont{K.}~\bibnamefont{Sekimoto}}, \bibinfo{journal}{Prog. Theor. Phys. Supp.} \textbf{\bibinfo{volume}{130}}, \bibinfo{pages}{17} (\bibinfo{year}{1998}), \eprint{doi:10.1143/PTPS.130.17}.

\bibitem[{\citenamefont{Halgren}(1996)}]{halgren1996merck}
\bibinfo{author}{\bibfnamefont{T.~A.} \bibnamefont{Halgren}}, \bibinfo{journal}{Journal of computational chemistry} \textbf{\bibinfo{volume}{17}}, \bibinfo{pages}{490} (\bibinfo{year}{1996}).

\bibitem[{\citenamefont{Sabhapandit}(2012{\natexlab{b}})}]{sabhapandit2012heat}
\bibinfo{author}{\bibfnamefont{S.}~\bibnamefont{Sabhapandit}}, \bibinfo{journal}{Physical Review E} \textbf{\bibinfo{volume}{85}}, \bibinfo{pages}{021108} (\bibinfo{year}{2012}{\natexlab{b}}).

\bibitem[{\citenamefont{Craven et~al.}(2018)\citenamefont{Craven, Chen, and Nitzan}}]{craven18a1}
\bibinfo{author}{\bibfnamefont{G.~T.} \bibnamefont{Craven}}, \bibinfo{author}{\bibfnamefont{R.}~\bibnamefont{Chen}}, \bibnamefont{and} \bibinfo{author}{\bibfnamefont{A.}~\bibnamefont{Nitzan}}, \bibinfo{journal}{J. Chem. Phys.} \textbf{\bibinfo{volume}{149}}, \bibinfo{pages}{104103} (\bibinfo{year}{2018}), \eprint{doi:10.1063/1.5007854}.

\bibitem[{\citenamefont{Craven and Nitzan}(2018)}]{craven18a2}
\bibinfo{author}{\bibfnamefont{G.~T.} \bibnamefont{Craven}} \bibnamefont{and} \bibinfo{author}{\bibfnamefont{A.}~\bibnamefont{Nitzan}}, \bibinfo{journal}{J. Chem. Phys.} \textbf{\bibinfo{volume}{148}}, \bibinfo{pages}{044101} (\bibinfo{year}{2018}), \eprint{doi:10.1063/1.5007854}.

\bibitem[{\citenamefont{Chua}(2014)}]{chua2014if}
\bibinfo{author}{\bibfnamefont{L.}~\bibnamefont{Chua}}, \bibinfo{journal}{Semiconductor Science and Technology} \textbf{\bibinfo{volume}{29}}, \bibinfo{pages}{104001} (\bibinfo{year}{2014}).

\bibitem[{\citenamefont{Isaeva et~al.}(2019)\citenamefont{Isaeva, Barbalinardo, Donadio, and Baroni}}]{Isaeva2019modeling}
\bibinfo{author}{\bibfnamefont{L.}~\bibnamefont{Isaeva}}, \bibinfo{author}{\bibfnamefont{G.}~\bibnamefont{Barbalinardo}}, \bibinfo{author}{\bibfnamefont{D.}~\bibnamefont{Donadio}}, \bibnamefont{and} \bibinfo{author}{\bibfnamefont{S.}~\bibnamefont{Baroni}}, \bibinfo{journal}{Nat. Commun.} \textbf{\bibinfo{volume}{10}}, \bibinfo{pages}{3853} (\bibinfo{year}{2019}), \eprint{doi:10.1038/s41467-019-11572-4}.

\bibitem[{\citenamefont{Lepri et~al.}(2003)\citenamefont{Lepri, Livi, and Politi}}]{Lepri2003}
\bibinfo{author}{\bibfnamefont{S.}~\bibnamefont{Lepri}}, \bibinfo{author}{\bibfnamefont{R.}~\bibnamefont{Livi}}, \bibnamefont{and} \bibinfo{author}{\bibfnamefont{A.}~\bibnamefont{Politi}}, \bibinfo{journal}{Phys. Rep.} \textbf{\bibinfo{volume}{377}}, \bibinfo{pages}{1} (\bibinfo{year}{2003}), ISSN \bibinfo{issn}{0370-1573}, \eprint{doi:10.1016/S0370-1573(02)00558-6}.

\bibitem[{\citenamefont{Kalantar et~al.}(2021)\citenamefont{Kalantar, Agarwalla, and Segal}}]{Segal2021}
\bibinfo{author}{\bibfnamefont{N.}~\bibnamefont{Kalantar}}, \bibinfo{author}{\bibfnamefont{B.~K.} \bibnamefont{Agarwalla}}, \bibnamefont{and} \bibinfo{author}{\bibfnamefont{D.}~\bibnamefont{Segal}}, \bibinfo{journal}{Phys. Rev. E} \textbf{\bibinfo{volume}{103}}, \bibinfo{pages}{052130} (\bibinfo{year}{2021}), \eprint{doi:10.1103/PhysRevE.103.052130}.

\bibitem[{\citenamefont{Segal et~al.}(2003{\natexlab{b}})\citenamefont{Segal, Nitzan, and H{\"a}nggi}}]{segal2003thermal}
\bibinfo{author}{\bibfnamefont{D.}~\bibnamefont{Segal}}, \bibinfo{author}{\bibfnamefont{A.}~\bibnamefont{Nitzan}}, \bibnamefont{and} \bibinfo{author}{\bibfnamefont{P.}~\bibnamefont{H{\"a}nggi}}, \bibinfo{journal}{J. Chem. Phys.} \textbf{\bibinfo{volume}{119}}, \bibinfo{pages}{6840} (\bibinfo{year}{2003}{\natexlab{b}}), \eprint{doi:10.1063/1.1603211}.

\bibitem[{\citenamefont{Nicolin and Segal}(2010)}]{Simine2010FPUT}
\bibinfo{author}{\bibfnamefont{L.}~\bibnamefont{Nicolin}} \bibnamefont{and} \bibinfo{author}{\bibfnamefont{D.}~\bibnamefont{Segal}}, \bibinfo{journal}{Phys. Rev. E} \textbf{\bibinfo{volume}{81}}, \bibinfo{pages}{040102} (\bibinfo{year}{2010}), \eprint{doi:10.1103/PhysRevE.81.040102}.

\bibitem[{\citenamefont{Campbell et~al.}(2005)\citenamefont{Campbell, Rosenau, and Zaslavsky}}]{CampbellFPUT2005}
\bibinfo{author}{\bibfnamefont{D.~K.} \bibnamefont{Campbell}}, \bibinfo{author}{\bibfnamefont{P.}~\bibnamefont{Rosenau}}, \bibnamefont{and} \bibinfo{author}{\bibfnamefont{G.~M.} \bibnamefont{Zaslavsky}}, \bibinfo{journal}{Chaos} \textbf{\bibinfo{volume}{15}}, \bibinfo{pages}{015101} (\bibinfo{year}{2005}), ISSN \bibinfo{issn}{1054-1500}, \eprint{doi:10.1063/1.1889345}.

\bibitem[{\citenamefont{Dhar and Roy}(2006)}]{dhar2006heat}
\bibinfo{author}{\bibfnamefont{A.}~\bibnamefont{Dhar}} \bibnamefont{and} \bibinfo{author}{\bibfnamefont{D.}~\bibnamefont{Roy}}, \bibinfo{journal}{J. Stat. Phys.} \textbf{\bibinfo{volume}{125}}, \bibinfo{pages}{801} (\bibinfo{year}{2006}), \eprint{doi:10.1007/s10955-006-9235-3}.

\bibitem[{\citenamefont{Portugal et~al.}(2024)\citenamefont{Portugal, Brange, and Flindt}}]{Portugal2024prl}
\bibinfo{author}{\bibfnamefont{P.}~\bibnamefont{Portugal}}, \bibinfo{author}{\bibfnamefont{F.}~\bibnamefont{Brange}}, \bibnamefont{and} \bibinfo{author}{\bibfnamefont{C.}~\bibnamefont{Flindt}}, \bibinfo{journal}{Phys. Rev. Lett.} \textbf{\bibinfo{volume}{132}}, \bibinfo{pages}{256301} (\bibinfo{year}{2024}), \urlprefix\url{https://link.aps.org/doi/10.1103/PhysRevLett.132.256301}.

\end{thebibliography}

\end{document}